\documentclass[useAMS,usenatbib]{mn2e}  

\usepackage{dsfont}
\usepackage{amsmath}
\usepackage{amssymb}
\usepackage{graphicx}
\usepackage{verbatim}
\usepackage{natbib}
\usepackage{eucal}
\usepackage{calligra}

\usepackage{amsfonts}
\usepackage{color}
\usepackage[normalem]{ulem}
\usepackage[T1]{fontenc}

\usepackage{times,epstopdf}
\usepackage{epsfig}
\usepackage[usenames,dvipsnames]{xcolor}
\usepackage{url}
\usepackage{subfigure}

\DeclareMathAlphabet{\mathscr}{OT1}{pzc}%
                                 {m}{it}

\newcommand{\be}{\begin{equation}}
\newcommand{\ee}{\end{equation}}
\newcommand{\bes}{\begin{equation*}}
\newcommand{\ees}{\end{equation*}}
\newcommand{\bea}{\begin{eqnarray}}
\newcommand{\eea}{\end{eqnarray}}
\newcommand{\beas}{\begin{eqnarray*}}
\newcommand{\eeas}{\end{eqnarray*}}

\newcommand{\mpl}{M_{\rm Pl}}

\newcommand{\tcr}{\textcolor{black}}

\begin{document} 

\title[Testing Gravity using Cosmic Voids]
  {Testing Gravity using Cosmic Voids}
\author[Cai et al.]
 {Yan-Chuan~Cai$^{1}$\thanks{E-mail: y.c.cai@durham.ac.uk}, Nelson Padilla $^{2,3}$\thanks{E-mail: npadilla@astro.puc.cl}, Baojiu~Li$^{1,}$\thanks{E-mail: baojiu.li@durham.ac.uk}\\
  $^1$Institute for Computational Cosmology, Department of Physics, University of Durham, South Road, Durham DH1 3LE, UK \\
$^2$Instituto de Astrof\'isica, Pontificia Universidad Cat\'olica, Av. Vicu\~na Mackenna 4860, Santiago, Chile. \\
$^3$Centro de Astro-Ingenier\'ia, Pontificia Universidad Cat\'olica, Av. Vicu\~na Mackenna 4860, Santiago, Chile.
}
\maketitle

\begin{abstract}
We explore voids in dark matter and halo fields from simulations of $\Lambda$CDM and Hu-Sawicki $f(R)$ models. In $f(R)$ gravity, dark matter void abundances are greater than that of general relativity (GR). However, when using haloes to identify voids, the differences of void abundances become much smaller, but can still be told apart, in principle, at the $2, 6$ and $14 \sigma$ level for the $f(R)$ model parameter amplitudes of $|f_{R0}|=10^{-6}$, $10^{-5}$ and $10^{-4}$. In contrast to the naive expectation, the abundance of large voids found using haloes in $f(R)$ gravity is lower than in GR. The more efficient halo formation in underdense regions makes $f(R)$ voids less empty of haloes. This counter intuitive result suggests that voids are not necessarily emptier in $f(R)$ if one looks at galaxies or groups in voids. Indeed, the halo number density profiles of voids are not distinguishable from GR. However, the same $f(R)$ voids are more empty of dark matter. This can in principle be observed by weak gravitational lensing of voids, for which the combination of a spec-$z$ and a photo-$z$ survey over the same sky is necessary. For a volume of  1~(Gpc/$h$)$^3$, neglecting the lensing shape noise,  $|f_{R0}|=10^{-5}$ and $10^{-4}$ may be distinguished from GR using the lensing tangential shear signal around voids by 4 and 8$\sigma$.  We find that the line-of-sight projection of large-scale structure is the main systematics that limits the significance of this signal, limiting the constraining power for $|f_{R0}|=10^{-6}$. We expect that this can be overcome with larger volume. The halo void abundance being smaller and the steepening of dark matter void profiles in $f(R)$ models are unique features that can be combined to break the degeneracy between $|f_{R0}|$ and $\sigma_8$. The outflow of mass from void centers are 5\%, 15\% and 35\% faster in $f(R)$ for $|f_{R0}|=10^{-6}$, $10^{-5}$ and $10^{-4}$, with little dependence on the tracers used. The velocity dispersions are greater than that in GR by similar amounts.  The absolute differences are greater for large voids. Model differences in velocity profiles imply potential powerful constraints of the model in phase space and in redshift space. 
\end{abstract}


\section{Introduction}

Cosmic voids as cosmological probes have been explored in many aspects.
For example, the Alcock-Paczynski test using stacking of voids has been demonstrated to be a powerful probe of distance 
distortions between the line-of-sight and transverse directions
\citep{Lavaux2012, SutterAP2014}; stacking of voids for the cosmic microwave background has been used {as an alternative 
to the cross-correlation method} to detect the integrated Sachs-Wolfe effect \citep{CaifR2014,Cai2013,Granett2008,Hotchkiss2014,Ilic2013,Planck2013}; 
the weak gravitational lensing effect of voids, as an analogue to that of haloes, has been shown to be capable of measuring 
the matter density profiles of voids \citep{Clampitt2014,Melchior2014}; void ellipticity has been shown to be sensitive to 
the dark energy equation of state \citep{Bos2012}; void properties have been studied in dynamical dark energy \citep{vdW2012}, 
coupled dark energy \citep{LiVoid2011,Sutter2014} and modified gravity \citep{lzk2012} models using N-body simulations. 

The basic properties of voids -- their abundances and density profiles -- can be powerful in constraining theories of
cosmology and gravity, but there are still gaps between observations and theoretical predictions for these two 
properties. The reasons are partly related to the technical details of how voids are defined: 
1) given the same simulation, different void-finding algorithms do not usually 
find the same voids \citep{Colberg2008}; 2) voids defined using tracers (galaxies or haloes) are expected to be different from voids  {defined using the dark matter field}. 
Nevertheless, studies of void profiles have made encouraging progress in the last decades.  Applying a 
spherical {under}-density algorithm to find voids in N-body simulations, \citet{Colberg2005} have found that void
profiles are self-similar within the effective void radius they defined, \citep[but see][]{Ricciardelli2014}. 
Using {\sc Zobov} \citep{Neyrinck05}, \citet{Hamaus2014} have found that dark matter void profiles 
within a wide range of  {radii} can be similar if rescaled using two free parameters. Using mocks and real LRG galaxies from SDSS, 
\citet{Nadathur2014} found an even simpler rescaling relationship among voids of different sizes, 
using only one parameter. 

It is perhaps more challenging to use void abundance to constrain cosmology. Excursion set theory offers 
predictions of the void abundance but the agreement with N-body simulations is never as great as in the case of haloes 
\citep{Sheth2004,Colberg2005,Paranjape2011,Achitouv2013}. A recent attempt to compare void abundance between simulations and 
excursion set theory has found that they may agree at $\sim16\%$ at $z=0$ if dark matter is used to define voids in 
simulations \citep{Jennings2013}, but the agreement is much worse if haloes are used instead. It would be even harder 
to compare theoretical predictions with observed void abundances from galaxy surveys due to the systematics  {introduced by}
survey geometry and masks. One possible way to overcome this is to apply the same algorithm to simulations 
and observations, which makes comparisons of void properties with simulations meaningful.

In this work, we will focus on {the potential of} using the above two basic properties of voids to constrain modified gravity.
Scalar-field models of modified gravity could drive the late-time accelerated expansion of the Universe without {explicitly} invoking a
cosmological constant. {Although such theories often introduce extra long-range forces (known 
as fifth forces) mediated by the scalar fields, they can still in principle pass} local tests of gravity via certain screening mechanisms, 
such as the Vainshtein \citep{Vainshtein1972} or chameleon \citep{kw2004} 
mechanisms. More explicitly, the success of those models largely relies on the screening mechanism to suppress the fifth force
in over-dense regions, where most of current astronomical observables come from. It is therefore inbuilt in these models that 
their differences from general relativity (GR) are minimal in high density environments like dark matter haloes, 
galaxies and the local Solar system. Nevertheless, in these models, the growth of structure is altered to some extent 
due to the coupling of scalar fields with matter, though the ensemble average of the growth at large scales might still be 
the same as in ${\rm \Lambda}$CDM \citep{Jennings2012,LiPS2013}. For example, in the $f(R)$ model \citep{hs2007}, 
which is mathematically equivalent to the coupled scalar field model mentioned above, the difference from ${\rm \Lambda}$CDM in terms of the 
halo mass function and matter power spectrum are not yet in obvious contradiction with observational data (see below). It is 
therefore interesting and necessary to explore alternative probes that are sensitive to the environment-dependent 
nature of these models. 

Voids in modified gravity have previously been studied by \citet{lzk2012} using N-body simulations, and 
by \citet{Martino2009,Clampitt2013,Lam2014} using analytical methods. {For chameleon-type} coupled scalar field models, 
the fifth force is found to counter the standard Newtonian gravity in under-densities \citep{Clampitt2013}. The repulsive force drives under-densities to 
expand faster and grow larger, hence also changing the distribution and abundances of voids. 
This naturally {implies that the two basic properties of voids, their density profiles and abundances, can be promising tools to probe and constrain modified gravity.} 

In this work, we use a set of N-body simulations of the $f(R)$ model to study properties of voids in detail, focusing on 
prospective observables to distinguish the $f(R)$ model from GR using voids. In section 2, the basics of the $f(R)$ models and the N-body simulations 
are introduced. Section 3 summarises our void-finding algorithm. We present the main results of comparing void properties, and discuss the 
observational implications in Section 4 and Section 5.



\section{The $f(R)$ gravity model and its simulations}

In this section we briefly introduce the theory and simulations of $f(R)$ gravity, to make this paper self contained.

\subsection{The $f(R)$ gravity model}

The $f(R)$ gravity model \citep{cddett2005} generalises GR by simply replacing the Ricci scalar $R$ in the standard Einstein-Hilbert action by an algebraic function 
$f(R)$~\citep[see e.g.][for some recent reviews]{sf2010, dt2010}:
\begin{eqnarray}\label{eq:fr_action}
S = \int{\rm d}^4x\sqrt{-g}\left\{\frac{\mpl^2}{2}\left[R+f(R)\right]+\mathcal{L}_m(\psi_i)\right\},
\end{eqnarray}
where $\mpl$ is the reduced Planck mass, and is related to Newton's constant $G$ by $\mpl^{-2}=8\pi G$, $g$ is the determinant of the metric tensor $g_{\mu\nu}$ and $\mathcal{L}_m(\psi_i)$
the Lagrangian density for matter, in which $\psi_i$ symbolically denotes the matter field for the $i$-th species, $i$ running over photons, neutrinos, baryons and cold dark matter. By fixing the functional form of $f(R)$, one fully specifies a $f(R)$ gravity model.

Variation of the action Eq.~(\ref{eq:fr_action}) with respect to the metric tensor $g_{\mu\nu}$ leads to the modified Einstein equation
\begin{eqnarray}\label{eq:fr_einstein}
G_{\mu\nu} + f_RR_{\mu\nu} -g_{\mu\nu}\left[\frac{1}{2}f-\Box f_R\right]-\nabla_\mu\nabla_\nu f_R = 8\pi GT^m_{\mu\nu},
\end{eqnarray}
where $G_{\mu\nu}\equiv R_{\mu\nu}-\frac{1}{2}g_{\mu\nu}R$ is the Einstein tensor, $f_R\equiv {\rm d} f/{\rm d} R$, $\nabla_{\mu}$ is the covariant derivative
compatible to $g_{\mu\nu}$, $\Box\equiv\nabla^\alpha\nabla_\alpha$ and $T^m_{\mu\nu}$ the energy momentum tensor for matter fields. Eq.~(\ref{eq:fr_einstein}) is a differential equation containing up to fourth-order derivatives of the metric tensor, because $R$ itself contains second derivatives of $g_{\mu\nu}$. It is often helpful to consider it as the standard second-order Einstein equation for general relativity, which is additionally sourced by a new, scalar, dynamical degree of freedom -- the so-called scalaron field $f_R$.  The equation of motion for $f_R$ can be derived by simply taking the trace of Eq.~(\ref{eq:fr_einstein}), as
\begin{eqnarray}\label{eq:fr_eom}
\Box f_R = \frac{1}{3}\left(R-f_RR+2f+8\pi G\rho_m\right),
\end{eqnarray}
where $\rho_m$ is the matter density. Note that in writing this equation we have assumed no massive neutrinos. This equation does not assume that radiations (photons and massless neutrinos) are negligible, because both are traceless, but in what follows we do assume this, since we will be considering the late-time cosmology of $f(R)$ gravity.

We consider a universe with no spatial curvature, and thus the background evolution is described by the flat Friedman-Robertson-Walker (FRW) metric. The line element of the real, perturbed, Universe is expressed in the Newtonian gauge as
\begin{eqnarray}
{\rm d} s^2 = a^2(\eta)\left[(1+2\Phi){\rm d}\eta^2 - (1-2\Psi){\rm d} x^i{\rm d} x_i\right],
\end{eqnarray}
where $\eta$ and $x^i$ are the conformal time and comoving coordinates, and $\Phi(\eta,{\bf x})$, $\Psi(\eta,{\bf x})$ are respectively the Newtonian
potential and perturbed spatial curvature, which depend on both time $\eta$ and space ${\bf x}$; $a$ is the scale factor of the
Universe and $a=1$ today.

We are interested in the formation of large-scale structures on scales substantially below the horizon. Under this assumption, the time variation of $f_R$ is small for the models to be studied, and thus we shall work with the quasi-static approximation by neglecting the time derivatives of $f_R$ in all field equations. The equation of motion of $f_R$, Eq.~(\ref{eq:fr_eom}), then 
reduces to
\begin{eqnarray}\label{eq:fr_eqn_static}
\vec{\nabla}^2f_R = -\frac{1}{3}a^2\left[R(f_R)-\bar{R} + 8\pi G\left(\rho_m-\bar{\rho}_m\right)\right],
\end{eqnarray}
in which $\vec{\nabla}$ is the 3-dimensional gradient operator (we use an arrow to distinguish this from the $\nabla_\mu$ introduced above), and the overbar takes
the background averaged value of the quantity under it. Note that $R$ has been expressed as a function of $f_R$ by reverting $f_R(R)$.

Also in the quasi-static approximation, the (modified) Poisson equation, which determines the behaviour of the Newtonian potential $\Phi$, can be written as
\begin{eqnarray}\label{eq:poisson_static}
\vec{\nabla}^2\Phi = \frac{16\pi G}{3}a^2\left(\rho_m-\bar{\rho}_m\right) + \frac{1}{6}a^2\left[R\left(f_R\right)-\bar{R}\right],
\end{eqnarray}
where, in addition to the time derivatives of $f_R$, we have also neglected those of $\Phi$. We have used Eq.~(\ref{eq:fr_eqn_static}) to eliminate $\vec{\nabla}^2f_R$ from Eq.~(\ref{eq:poisson_static}).

From the above equations, we see that there are two potential cosmological implications of the scalaron field: (i) the background expansion of the Universe can
be altered by the new terms in Eq.~(\ref{eq:fr_einstein}) and (ii) the relationship between the gravitational potential $\Phi$ and the matter density perturbation is altered, which
can cause changes to the growth of density perturbations. 

It is useful to inspect in which conditions the general relativistic (or Newtonian) limit is recovered. Evidently, when $|f_R|\ll1$, Eq.~(\ref{eq:fr_eqn_static}) gives $R\approx-8\pi G\rho_m$ 
and so Eq.~(\ref{eq:poisson_static}) reduces to the normal Poisson equation of general relativity: 
\begin{eqnarray}
\vec{\nabla}^2\Phi = 4\pi{G}a^2\left(\rho_m-\bar{\rho}_m\right).
\end{eqnarray}
When $|f_R|$ is large, we instead
have $|R-\bar{R}|\ll8\pi G|\rho_m-\bar{\rho}_m|$ and then Eq.~(\ref{eq:poisson_static}) reduces to the normal Poisson equation but with $G$ enhanced 
by $1/3$:
\begin{eqnarray}
\vec{\nabla}^2\Phi = \frac{16\pi{G}}{3}a^2\left(\rho_m-\bar{\rho}_m\right).
\end{eqnarray}
The value $1/3$ is the maximum enhancement factor of the strength of gravitational interaction in $f(R)$ models, independent of the specific functional form of $f(R)$.
Nevertheless, the choice of $f(R)$ is important because it determines the scalaron dynamics and therefore at what time and scale the enhancement factor switches from $1/3$ to $0$: scales substantially larger than the Compton wavelength of $f_R$ (which characterises the range of the scalaron-mediated modification to Newtonian gravity) are unaffected and gravity is as predicted by general relativity there; on small scales, however, depending on the the dynamical properties of Eq.~(\ref{eq:fr_eqn_static}), the $1/3$ enhancement could be fully realised -- this leads to a scale-dependent modification of gravity and consequently scale-dependent
evolution of cosmic structures.

\subsection{The chameleon mechanism}

\label{subsect:fr_cham}

The $f(R)$ model would have been ruled out by Solar system gravity tests due to the factor-of-$4/3$ rescaling to the strength of Newtonian gravity.
Nevertheless, it is known that, if the functional form of $f(R)$ is chosen carefully \citep{bbh2006,nv2007,lb2007,hs2007,bbds2008}, the model can use
the so-called chameleon mechanism \citep{kw2004} to suppress the effects of the scalaron field and therefore reduce to standard gravity in regions with deep enough gravitational potentials, such as our Solar system.

The chameleon mechanism can be understood as the following: the modification to the Newtonian gravity is usually considered as an extra, or fifth, force
mediated by the scalaron field. Because the scalaron, unlike gravitons of general relativity, has a mass, this extra force has the Yukawa form, which decays exponentially 
when the distance $r$ between two test masses increases, as $\propto\exp(-mr)$, in which $m$ is the scalaron mass, defined as
\begin{eqnarray}
m^2 \equiv \frac{{\rm d}^2V_{\rm eff}\left(f_R\right)}{{\rm d}f_R^2},
\end{eqnarray}
in which $V_{\rm eff}$ is the effective potential for the scalaron (see below), which usually has a minimum which depends on local matter density. In high matter density environments, $m$ is very
heavy and the exponential decay causes a strong suppression of the fifth force. In terms of the effective potential, this is equivalent to saying that $f_R$ is trapped to the minimum of $V_{\rm eff}(f_R)$, which itself is restricted to be very close to zero: consequently, $|f_R|\ll1$ in such environments, and this leads to the GR limit as we have discussed above\footnote{Although it is not technically incorrect to say that the chameleon mechanism works in high-density regions, those regions need to be large enough, e.g., at least comparable to the local Compton wavelength of the scalaron field. Otherwise, the chameleon mechanism does not necessarily work. For this reason, it is more accurate to say that the screening happens in regions with deep Newtonian potential. Nevertheless, in this paper we shall not distinguish between the two.}.

As we will see below, $V_{\rm eff}$ is determined by the functional form of $f(R)$. Consequently, as mentioned earlier, the latter plays a crucial role in deciding whether the fifth force can be sufficiently suppressed in dense
environments. In this paper we will focus on the $f(R)$ model proposed by \citet{hs2007}, which is specified by
\begin{eqnarray}\label{eq:hs}
f(R) = -M^2\frac{c_1\left(-R/M^2\right)^n}{c_2\left(-R/M^2\right)^n+1},
\end{eqnarray}
where $M^2\equiv8\pi G\bar{\rho}_{m0}/3=H_0^2\Omega_m$ is a new mass scale, with $H$ being the Hubble expansion rate and $\Omega_m$ the present-day fractional energy
density of matter. Hereafter, a subscript $_0$ always means taking the present-day ($a=1$) value of a quantity. 

At the cosmological background level, the scalaron field $f_R$ always closely follows the minimum of $V_{\rm eff}$, which itself is related to $f(R)$ (and local matter density - remember the chameleon mechanism depends on this) by 
\begin{eqnarray}
V_{\rm eff}\left(f_R\right) \equiv \frac{1}{3}\left(R-f_RR+2f+8\pi G\rho_m\right).
\end{eqnarray}
Therefore we find
\begin{eqnarray}
-\bar{R} \approx 8\pi G\bar{\rho}_m-2\bar{f} \approx 3M^2\left(a^{-3}+\frac{2c_1}{3c_2}\right),
\end{eqnarray}
where for the first `$\approx$' we have used $|f_R|\ll1$, which holds true for the models to be studied in this paper (see below), and for the second we have used the fact that $|\bar{R}|\gg M^2$ throughout the evolution history of our Universe (see the discussion of Eq.~(\ref{eq:temp}) below).

To reproduce the background expansion history of the standard $\Lambda$CDM paradigm, we set
\begin{eqnarray}
\frac{c_1}{c_2} = 6\frac{\Omega_\Lambda}{\Omega_m}
\end{eqnarray}
where $\Omega_\Lambda=1-\Omega_m$ is the present-day fractional energy densities of `dark energy' (remember that we have neglected radiations at late times).

By setting $\Omega_\Lambda=0.76$ and $\Omega_m=0.24$\footnote{These values are currently outdated, but they have been used in the $f(R)$ simulations extensively in the literature, including the ones on which this paper is based. We remark that for the purpose of this study, the exact value of $\Omega_m$ is not crucial, as confirmed by some test simulations with different $\Omega_m$.}, we find that $|\bar{R}|\approx41M^2\gg M^2$ today (and $|\bar{R}|$ is even larger at early times), and
this simplifies the expression of the scalaron to
\begin{eqnarray}\label{eq:temp}
f_R \approx -n\frac{c_1}{c_2^2}\left(\frac{M^2}{-R}\right)^{n+1} < 0.
\end{eqnarray}
Therefore, the two free parameters $n$ and $c_1/c_2^2$ completely specify this special $f(R)$ model. Moreover, they are related to the value of 
the scalaron today ($f_{R0}$) by
\begin{eqnarray}
\frac{c_1}{c_2^2} = -\frac{1}{n}\left[3\left(1+4\frac{\Omega_\Lambda}{\Omega_m}\right)\right]^{n+1}f_{R0}.
\end{eqnarray}
In the present paper we will study three $f(R)$ models with $n=1$ and $|f_{R0}|=10^{-6}, 10^{-5}, 10^{-4}$, which will be referred to as F6, F5 and F4 
respectively. These choices of $|f_{R0}|$ are designed to cover the part of the parameter space which is cosmologically interesting: if $|f_{R0}|>10^{-4}$
then the $f(R)$ model violates the cluster abundance constraints \citep{svh2009}, and if $|f_{R0}|<10^{-6}$ then the model is too close to
$\Lambda$CDM to be distinguishable in cosmology. \cite{lombriser2014} gives an excellent review on the current cosmological and astrophysical constraints on $f_{R0}$.

The environmental dependence of the screening means that gravity can behave very differently in different situations. The matter clustering in overdense regions, e.g., dark matter haloes, in $f(R)$ gravity has been studied in great detail, and research shows that the chameleon mechanism can be very efficient in suppressing the fifth force. We will focus on voids, i.e., underdense regions, in this paper, where the chameleon mechanism is expected to work less well and so deviation from standard Newtonian gravity would be stronger. 

\subsection{The N-body simulations of $f(R)$ gravity}

According to Eqs.~(\ref{eq:fr_eqn_static}, \ref{eq:poisson_static}), if the matter density field is known, one can solve the scalaron field $f_R$
from Eq.~(\ref{eq:fr_eqn_static}) and plug the solution into the modified Poisson equation, Eq.~(\ref{eq:poisson_static}), to compute $\Phi$. When $\Phi$ is
known, a numerical differentiation will give the total (modified) gravitational force that determines the motions of particles. Such is the basic logic and procedure of $f(R)$ $N$-body simulations.

The major challenge, then, is to numerically solve the highly nonlinear scalaron equation of motion, Eq.~(\ref{eq:fr_eqn_static}), to accurately calculate the gravitational force when the chameleon screening is working. Since there is no analytical force law (such as the $r^{-2}$-law in Newtonian gravity), it is usually convenient to solve $f_R$ on a mesh (or a suite of meshes) using relaxation methods. This implies that mesh-based $N$-body
codes are most suitable.

$N$-body simulations of $f(R)$ gravity or chameleon theories have previously been performed by \citet{oyaizu2008,lz2009,zlk2011}. Because the highly nonlinear nature of Eq.~(\ref{eq:fr_eqn_static}) means that the code spends a substantial fraction of the computation time solving $f_R$, early simulations were largely
limited by the box size or force resolution, or both. This work makes use of large-box $f(R)$ simulations performed using the {\sc ecosmog} code \citep{lztk2012}. {\sc ecosmog} is an extension to the mesh-based $N$-body code {\sc ramses} \citep{ramses}, which is parallelised with MPI and can therefore have improved efficiency and performance than early serial simulation codes for $f(R)$ gravity. The technical details of the code are described elsewhere, e.g., \cite{lztk2012} and interested readers are referred to those references. We note that parallelised $f(R)$ simulation codes have recently been developed by other groups, e.g., \cite{mg-gadget,isis}.

The simulations used for the void analyses in this work are listed in Table~\ref{table:simulations}. They all have the same values of cosmological parameters, summarised here as $\Omega_m = 0.24$, $\Omega_\Lambda = 0.76$, $h = 0.73$, $n_s = 0.958$ and $\sigma_8 = 0.80$, where $h\equiv H_0/(100$~km/s/Mpc) is the dimensionless Hubble parameter at present, $n_s$ is the spectral index of the scalar primordial power spectrum and $\sigma_8$ is the linear root-mean-squared density fluctuation measured in spheres of radius 8${\rm Mpc}/h$ at $z=0$. The background cosmology for all these models is the same as in the $\Lambda$CDM variant in practice (the difference caused by the different $f(R)$ model parameters is negligible). Therefore, the differences in these simulations only come from the modified gravitational law.

\begin{table*}
\caption{Some technical details of the simulations performed for this work. F6, F5 and F4 are the labels of the Hu-Sawicki $f(R)$ models with $n=1$ and
$|f_{R0}|=10^{-6}, 10^{-5}, 10^{-4}$ respectively. Here $k_{Nyq}$ denotes the Nyquist frequency. 
The last column lists the number of realisations for each simulation.}
\begin{tabular}{@{}lcccccc}
\hline\hline
models & $L_{\rm box}$ & number of particles & $k_{Nyq}$ $[h~\textrm{Mpc}^{-1}]$ & force resolution [$h^{-1}$kpc] & number of realisations \\
\hline
$\Lambda$CDM, F6, F5, F4 & $1.5h^{-1}$Gpc & $1024^3$ & 2.14 & 22.9 & $6$ \\
$\Lambda$CDM, F6, F5, F4 & $1.0h^{-1}$Gpc & $1024^3$ & 3.21 & 15.26 & $1$ \\
$\Lambda$CDM & 250${\rm Mpc}/h$ & $1024^3$ & 12.82 & 3.8 & 1\\
\hline
\end{tabular}
\label{table:simulations}
\end{table*}

We call a group of $\Lambda$CDM, F6, F5 and F4 simulations with the same technical settings (e.g., box size and resolution) a simulation suite. All models in each simulation suite share the same initial condition generated using the Zel'dovich approximation at an initial redshift, $z_i = 49$. Although generally speaking the modified gravitational law also affects the generation of the initial 
condition \citep{lb2011}, in the $f(R)$ models studied here the fifth force is strongly suppressed at $z_i\geq$ a few, and so its effects are negligible at the initial times. The use of the same initial conditions for all simulations in a suite means that the initial density fields for the $\Lambda$CDM and the $f(R)$ simulations share the same phases, and any difference in the void properties that we find at late times is a direct consequence of the 
different dynamics and clustering properties of $f(R)$ gravity.

\section{Void finding algorithm}

We use a void finder based on the one presented by \citet[P05]{Padilla2005} with improvements centred on
optimising the method, on improving its convergence for different mass resolutions and box sizes,
and adapted to run on parallel computers. Our finder can be run either on the dark matter
field or using dark matter halo tracers. 

In the case of searching for voids in the dark matter field,
the improved finder (iP05 from now on) follows the following steps:

\indent
(i) It searches throughout the simulation volume for regions of low
density, performing a top-hat smoothing of the density field on a grid.  We take all the grid cells with a density
below a set threshold $\delta_{\rm cell}$ as prospective void centres. 
We define a grid such that each cell contains on average $10$ dark-matter particles
{so as to avoid being affected by shot noise that unnecessarily increases the number of prospective centres 
when this minimum number is lower.
In the case when the detection of voids is done using haloes we adopt the same average number of objects in a cell (notice
that since haloes trace peaks {of the dark matter field}, this is probably a conservative choice).}
We adopt a grid threshold $\delta_{\rm cell}=-0.9$, i.e., only empty cells are prospective void centres.

\indent
(ii) We start measuring the underdensity in spheres
of increasing radius about these centres until some void threshold, $\Delta_{\rm void}$, is reached.
If the largest sphere about any one centre contains at least $20$ particles, it is kept as a void candidate.
{This number provides the best stability of the method against resolution and boxsize changes when searching
for voids in the dark matter density field.}
In almost all cases we adopt a value of $\Delta_{\rm void}=-0.8$.  However, for $z>0$ we also adopt different 
thresholds that take into account the linear growth of perturbations until $z=0$ (this will be indicated 
{for each particular case).}
The void radius is defined as the radius of the under-dense sphere.
{As a result, the void density profiles typically reach the cosmic mean density quite 
farther from the void radii}.

\indent
(iii) The candidates are ranked in decreasing order of size, and we reject all spheres which overlap with a
neighbouring sphere with a larger radius by more than a given percentage of the sum of their radii.  This
percentage will be treated as a free parameter: increasing the overlap percentage increases
the size of the sample, which helps to obtain better statistics, but it also introduces larger 
covariances in the results, which could in the end lower the significance of a comparison
between models.  In the case of voids in the distribution of the dark matter field, we choose to use $10$
percent overlap. 

Following Ceccarelli et al. (2012), 
voids are classified into void-in-void and void-in-cloud objects depending on whether the accumulated
overdensity at $3$ void radii from the void centre is below or above zero, respectively, {or simply by
dividing our sample in different void sizes, since smaller voids contain larger fractions of void-in-cloud objects
\citep[see][for examples]{Ceccarelli2013, Cai2013, Hamaus2014}}.

{
When searching for voids in the distribution of dark matter haloes, { step (ii) above is slightly modified},
where we {now} do not demand a minimum number of haloes within a void.  This is due to the high significance of the peaks
marked by haloes, which implies that a volume empty of haloes is one with no peaks, and as such it is
of interest to us.  Furthermore, if for the same initial conditions the resolution of the simulation were improved,
the haloes of the lower resolution simulation would again be found, {though} the dark matter particle positions would indeed change
due to the change in the sampling of the smooth density field.  
\tcr{Additionally, we will allow the percentage of void overlapping to vary from $50$ to $0$, in order to search for the optimal 
sample to distinguish between GR and $f(R)$ models.}

Using haloes to identify voids is more practical observationally than using dark matter, but one has to face the challenge of sparse sampling. Especially, the centering of the spherical voids needs to be done more carefully since haloes only sparsely sample the density field.  In particular, since a sphere can be defined by four points on its surface,
a well-centred spherical void should contain four haloes, all living at a distance
of exactly one void radius from the void centre.  However, the precision of centering of our voids is of $0.3$Mpc$/h$, so we do expect
some dispersion in the distance to the first four closest haloes.  We assess the quality of our voids using this dispersion defined
as $$\sigma_4\equiv\sqrt{\left< \left(\frac{d_i-\left< d_i\right>_4}{\left< d_i\right>_4}\right)^2 \right>_4},$$
in which $d_i$ is the distance from the void centre to the $i$-th halo, and $\left< \right>_4$ denotes the average over the 4 closest haloes.  The typical 
value for this parameter in our void samples is $\sigma_4\simeq0.05$, as shown in the inset of Fig.~\ref{fig:dndrhaloes}.

When using haloes as tracers, our void finder 
identifies voids with no haloes, but also allows voids to contain one or more
haloes within its volume, as long as the threshold criterion, $\Delta_{\rm void}=-0.8$, is met. {Note that we have $\Delta_{\rm void}=\delta n/\bar{n}$ when referring to threshold criterion for halo-identified voids, where $n$ is the halo number density; in contrast, for voids identified using the dark matter field, the threshold is defined as $\Delta_{\rm void}=\delta\rho_m/\bar{\rho}_m$.}

There are several routes to come closer to voids as would be identified with real galaxy catalogues.  For
example, one could simply weight each dark matter halo by the halo occupation number that corresponds to its mass.
To do this, though, one needs the halo occupation distribution (HOD) parameters fit to observational data, and this
is currently only available for $\Lambda$CDM \citep[e.g.][]{Zehavi}.  In order to make a consistent comparison
between GR and $f(R)$ it would be necessary to first find the HOD parameters for the $f(R)$ cosmologies.  Another
possible avenue is to use semi-analytic galaxies, obtained by coupling a semi-analytic galaxy formation
model to the merger trees extracted from the GR and $f(R)$ simulations.  
{This approach is again not ideal because it assumes that the physics of galaxy formation is the same in both models, which may not be the case.}
Our present approach of using haloes as tracers is comparable to identifying voids in galaxy group catalogues, and as such
is already applicable to current surveys.
}

\section{Void abundance}

\begin{figure}
\scalebox{0.43}{
\includegraphics[angle=0]{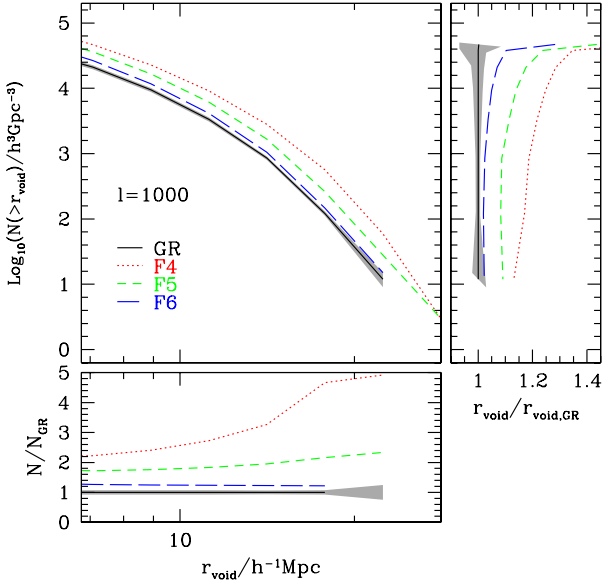}}
\vskip -0.75cm
\caption{Void abundance as a function of void radius in real space, 
for the GR, F6, F5, and F4 simulations of $1000$${\rm Mpc}/h$ a side boxes 
(different line types indicated in the key). The shaded region shows
the errors obtained from multiple realisations of a larger simulation.  The lower sub-panel
shows the abundance ratio between the different models and GR, while the right sub-panel shows the
ratios between void radii at fixed abundances with respect to GR.}
\label{fig:abundance}
\end{figure}

{We have discussed that voids can be identified using either the dark matter field or biased tracers such as dark matter haloes or galaxies. Although the latter are more directly related to real observational data, the former is of more theoretical interest, as it directly reflects how dark matter clusters under the action of modified gravitational laws. In this section, we show void abundances found in both ways.}

\subsection{Voids identified using the dark matter field}
 
We use the iP05 finder on the {dark matter field of the} $1000$~${\rm Mpc}/h$-aside boxes for the GR, F6, F5 and F4 simulations.
The resulting abundances as a function of void radius are shown in Figure~\ref{fig:abundance}.  We
show the results for the voids {identified using positions in real space for the dark-matter particles}.
In order to determine the minimum void radius above which our void sample is complete in these simulations,
we run the void finder in smaller $250$ ${\rm Mpc}/h$ simulations with the same number of particles, i.e.
with  {64} times the mass resolution, and find that the abundances of the small and large simulations
are consistent down to a void radius of $7$ ${\rm Mpc}/h$.  Therefore, from this point on, we will concentrate
on voids of at least this size {unless otherwise stated}.

As can be seen in the main panel of Figure~\ref{fig:abundance}, the abundance of voids at a fixed
radius increases from GR to $f(R)$ gravity, as expected due to the action of the fifth force that tends to point
towards the void walls \citep{Clampitt2013}, which is stronger progressing from F6 through F5 to F4. 
{Dark matter voids in $f(R)$ models therefore grow larger and their void abundances greater.}

The lower sub-panel shows the ratios of the abundances of the modified gravity models with respect
to GR. {The} F4 model shows the strongest differences from GR, with abundances higher by
factors of {$\sim1 - 4$}, and larger for larger void radii. The right sub-panel shows the ratio
of void sizes at fixed void number densities, where the roughly constant ratios indicate that the differences
between the models are almost due to larger void sizes for the modified gravity models {compared with} GR, 
except at high number densities where the ratios start to increase steadily.
{We checked the results in redshift space and found very small differences from what is shown in Fig.~\ref{fig:abundance}.}

\begin{figure}
\scalebox{0.59}{\includegraphics[angle=0]{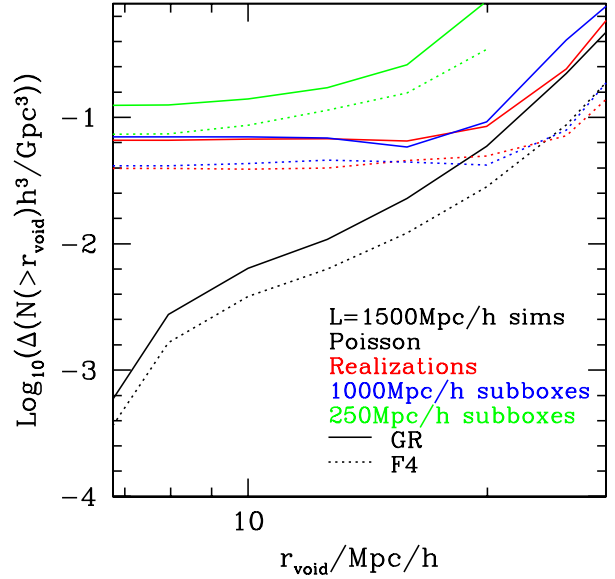}}
\vskip -3cm
\caption{Void abundance errors obtained from 6 realisations of our $1500$ ${\rm Mpc}/h$ a side simulations.
{We show errors arising from assuming Poisson fluctuations (black), the dispersion from the realisations for the full simulation
volumes (red), and for two smaller sub-volumes (blue and green)}, for the GR and F4 models (solid and dotted,
respectively).}
\label{fig:errors}
\end{figure}
 
The shaded region {in Fig.~\ref{fig:abundance}} shows the error in the GR abundances, and the lower panel 
shows that the GR and F6 abundances
are significantly different, by at least $3$ standard deviations ({3$\sigma$}) up to void radii $\simeq 20$ ${\rm Mpc}/h$. 
To estimate these, we have used larger simulations of
$1500$ ${\rm Mpc}/h$ a side, for which there are a total of $6$ realisations for each model.  In order to estimate
errors for the cubic Giga-parsec simulations, we simply compute the dispersion of abundances measured in 
$1000$ ${\rm Mpc}/h$-a side sub-boxes of the large simulations.  We use these errors instead of Poisson
estimates since the dispersion from multiple realisations is more representative of the cosmic variance. 
Fig.~\ref{fig:errors} shows different estimates of cumulative abundance errors for the GR (solid) and 
F4 (dotted lines) models for the large simulations. The black lines correspond to the Poisson estimates
of the errors and the  {red} lines to the dispersion of the different realisations using the full simulation
volume.  These two estimates of errors are very different for void radii $<20$ ${\rm Mpc}/h$; this is the
reason why we adopt the realisation dispersion. 
The other lines show the dispersion of $1000$ (blue) and $250$ ${\rm Mpc}/h$ (green)
aside sub-boxes.  As can be seen, decreasing the volume to $1~({\rm Gpc}/h)^3$ does not increase the errors as it would
be expected from Poisson fluctuations.  The errors for the smaller volume of $250$ ${\rm Mpc}/h$ are much larger, on the other hand.

\begin{figure}
\scalebox{0.43}{
\includegraphics[angle=0]{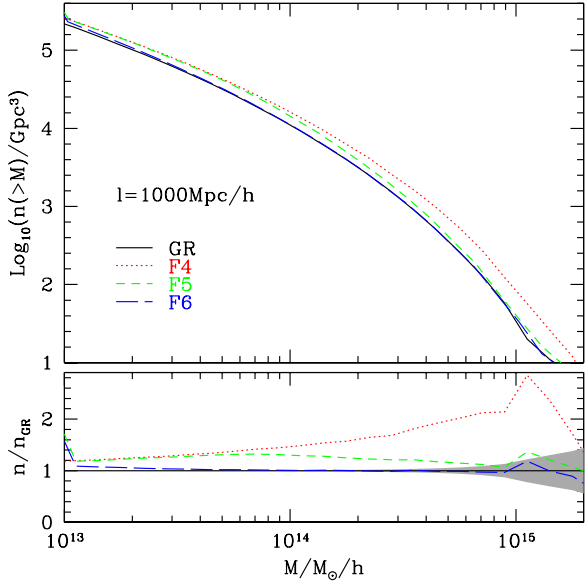}
}
\vskip -0.75cm
\caption{Cummulative halo mass functions for the GR, F6, F5, 
and F4 models (different line types and colours as indicated in the key). The shaded area shows the
Poisson error in the GR model.  The lower panel shows the ratio between the different
$f(R)$ models and GR, using the same colours as in the top panel.}
\label{fig:mfhalos}
\end{figure}

{As mentioned in the introduction, t}he main motivation to study voids in modified gravity models is that the abundances of voids are more
sensitive to the fifth force than the halo mass functions.  The latter is shown in Fig.~\ref{fig:mfhalos},
where it can be seen that the change of the halo abundance as we move from GR through F6 and F5 to F4 is
complicated.  Furthermore, the F6 model shows virtually no differences from the GR abundances
down to masses {$\simeq 10^{14}$ $h^{-1}M_{\odot}$}, whereas the void abundance is significantly different between
these two models across the range of void sizes reliably accessible in our simulations.
{The dark matter haloes are identified using the spherical overdensity code AHF \citep{ahf}.} 
 
\begin{figure}
\scalebox{0.43}{
\includegraphics[angle=0]{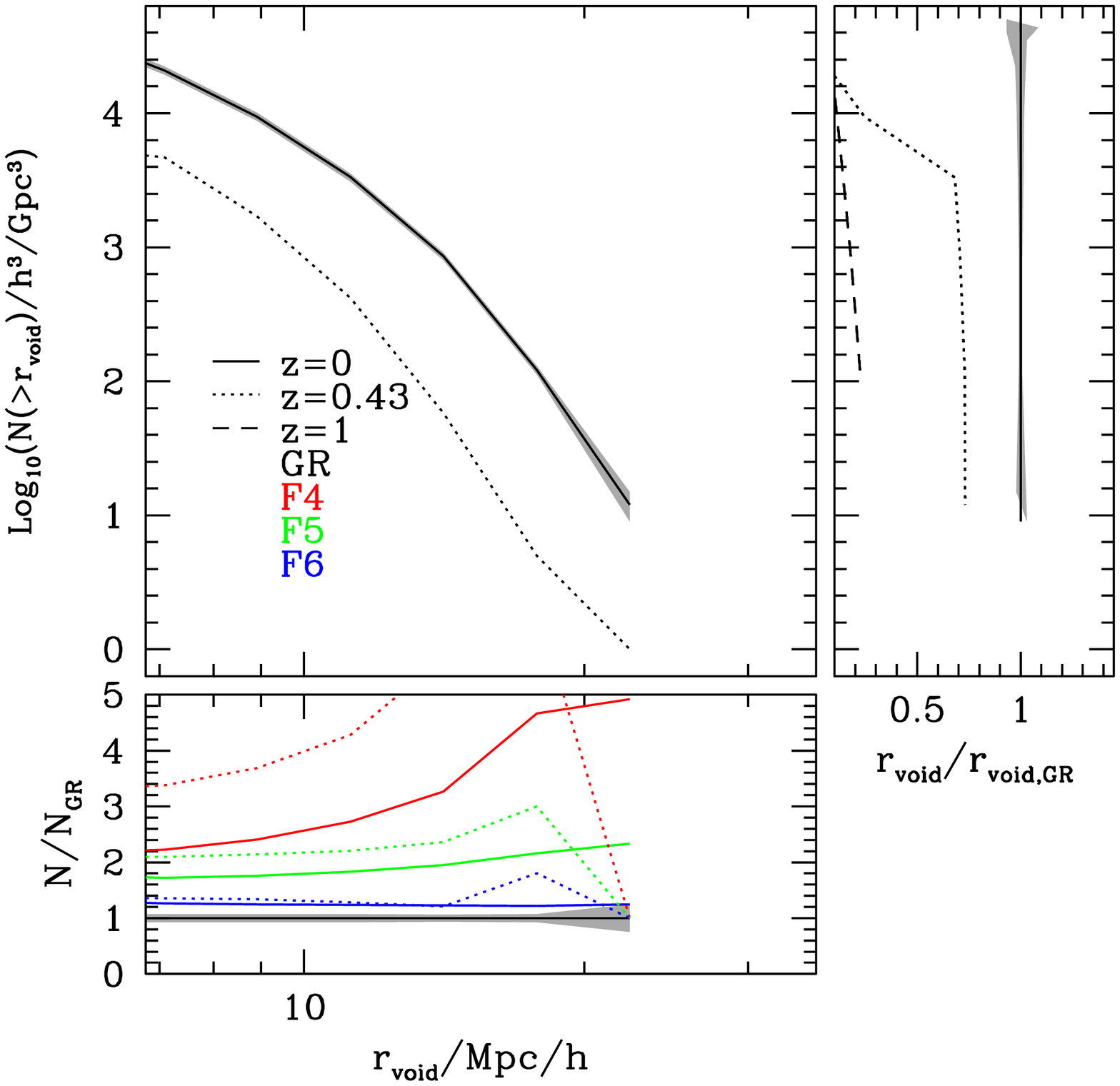}
}
\scalebox{0.43}{
\includegraphics[angle=0]{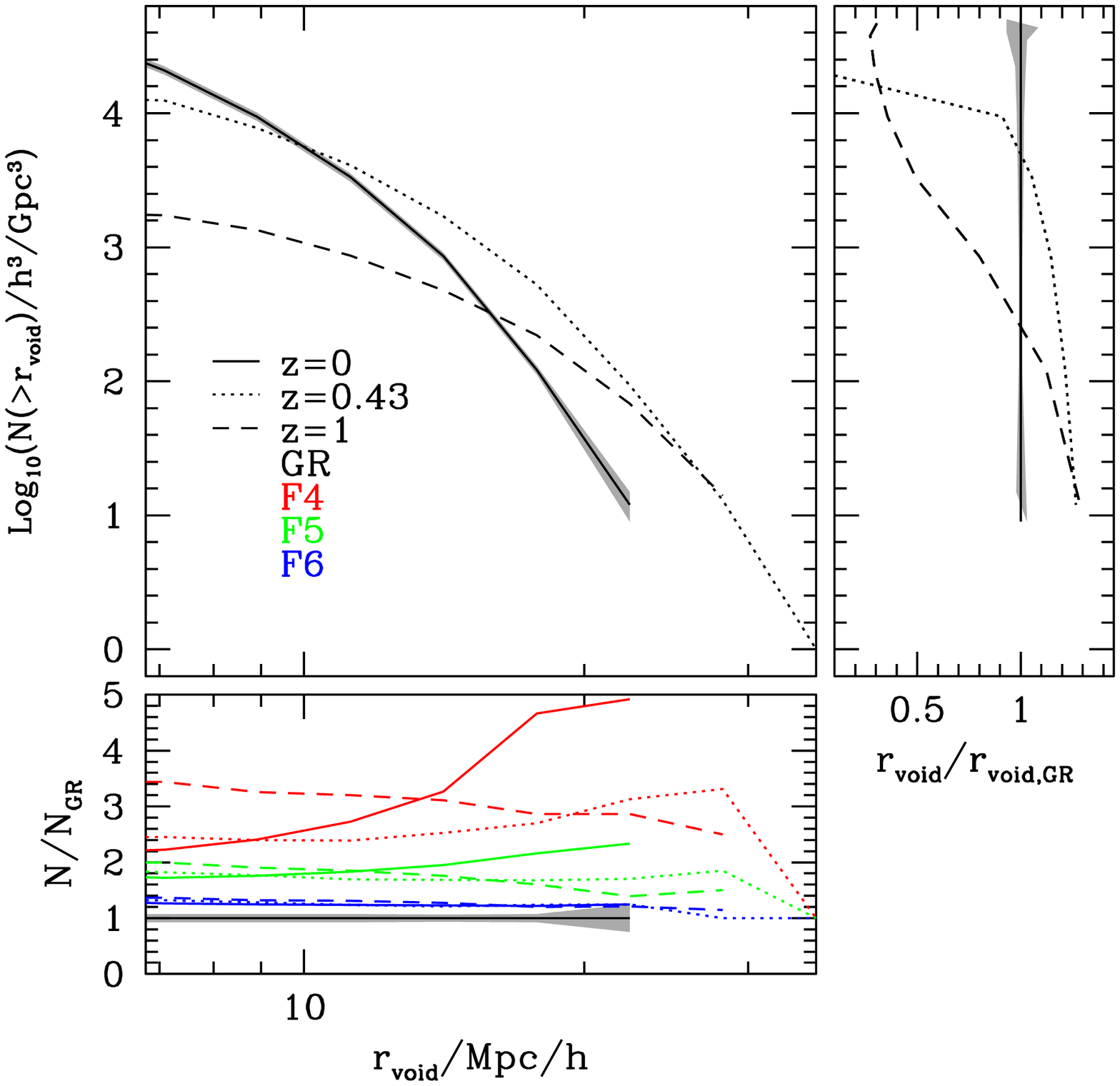}
}
\vskip -0.75cm
\caption{Void abundances at different redshifts {(solid, dotted and dashed as the redshift increases), }
adopting the same over-density threshold of
$\Delta_{\rm void}$ $=-0.8$ (top panels), and a threshold which evolves according to the linear perturbation growth, $\Delta_{\rm void}=-0.8, -0.65$ and $-0.51$ for $z=0,0.43$ and $1.0$, respectively (bottom
panels).  For clarity, the main panels only show the results for the GR simulation.
The lower sub-panels are as in Figure~\ref{fig:abundance}.  The right sub-panels show the ratio
between GR void radii at fixed abundance and redshift, with respect to GR voids at $z=0$. 
}
\label{fig:abundredshift}
\end{figure}

Given that large surveys are shifting toward higher redshifts,
we also measure the abundance of voids at {two} higher redshifts, $z=0.43$ and $1$.  Since
fluctuations grow with time one can either use a fixed density threshold to identify voids at different redshifts, or make it
evolve with the growth of fluctuations.  We adopt both approaches and show the results in
Fig.~\ref{fig:abundredshift}, where the top panels show the results for a fixed threshold of $\Delta_{\rm void}=-0.8$,
and the bottom panels for an evolving threshold following the linear growth of fluctuations for the
cosmology of the GR simulation (differences are almost indistinguishable for the modified gravity
models adopted here), with $\Delta_{\rm void}=-0.8, -0.65$ and $-0.51$ for $z=0,0.43$ and $1.0$, respectively.
To improve clarity the main panels only show the results for the GR simulations, but the lower
sub-panels show the ratio of abundances with respect to GR (colours for {different models, solid, dotted and dashed line types}
as the redshift increases).  The right sub-panels show the ratio between void sizes at fixed abundance
for GR alone, with respect to $z=0$.

The main panels of Figure~\ref{fig:abundredshift} show that a fixed threshold effectively produces much smaller samples at $z=0.43$ and $1$,
to the point that for the latter the number of voids is too small to appear in the figure. {This is because very empty regions, e.g., $\Delta_{\rm void}\leq-0.8$, only form at late times}. An evolving
threshold, on the other hand, produces higher abundances of large (comoving size) voids, but a smaller
total number of voids (the intersection of the lines with the $y$-axis). 
{This may be due to the use of linear theory to rescale the threshold, which is likely to be over-evolving the actual
growth of fluctuations because 
the evolution of under-densities tends to be slower than the linear theory result}. The right sub-panels show that 
the main difference in the abundance of voids for a fixed threshold can be absorbed by a constant factor in
the void radius: $z=0.43$ voids are $\simeq 0.7$ times smaller than those at $z=0$, whereas $z=1$ voids are
$\sim5$ times smaller than at $z=0$. The constant shifts of the void abundance functions among 
different epochs reflect a simple picture for the evolution of voids, which is captured by our void algorithm. 
More in-depth study of the evolution of void abundance is needed to shed light on the physics behind this behaviour.

Regarding the differences between $f(R)$ gravity and GR, at higher redshifts the abundances are 
higher in the F6 to F4 models with respect to GR, by factors similar to those at $z=0$ (see the lower
sub-panels).  However, if a fixed density threshold is adopted, the number of voids will be smaller and the errors
larger, making the difference between  F6 and GR less significant.  This should not be a problem if
the evolving threshold is adopted.
 
\begin{figure}
\scalebox{0.43}{\includegraphics[angle=0]{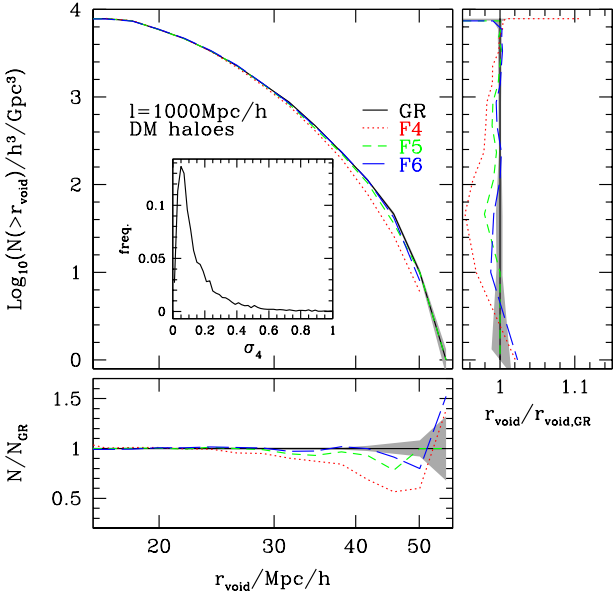}}
\vskip -0.75cm
\caption{{Void abundances resulting from using dark matter haloes as tracers of the density
field.  Panels and lines are as in Figure \ref{fig:abundance}}. {The inset in the main
panel shows the distribution function of $\sigma_4$, the relative dispersion in the distance
to the first four haloes above the void density threshold, for GR; a small value of $\sigma_4$ indicates
a well centred void.  } }
\label{fig:dndrhaloes}
\end{figure}

\begin{figure}
\scalebox{0.43}{\includegraphics[angle=0]{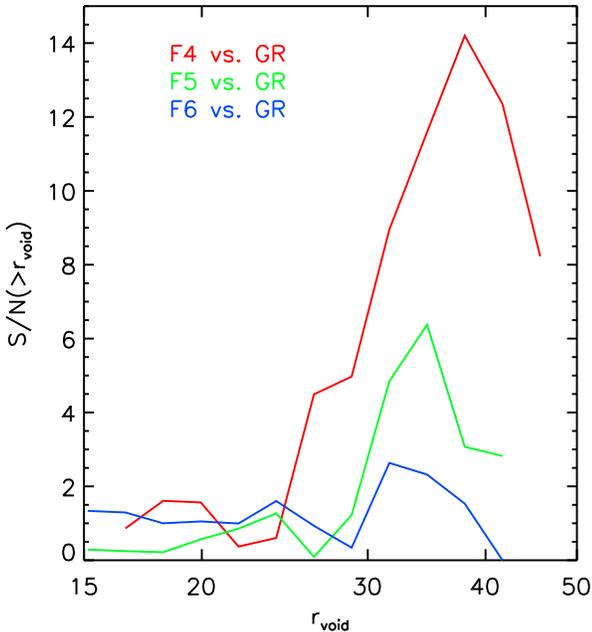}}
\caption{The corresponding S/N from void abundances shown in 
Figure \ref{fig:dndrhaloes}. }
\label{fig:dndrhaloes_SN}
\end{figure}

\subsection{{Voids identified using dark matter haloes}}

In real observations, voids are usually identified by using tracers of large-scale structure{, such as galaxies}. In general, to compare simulations with observational data, one has to generate galaxy mocks. {However, \citet{Padilla2005} has shown that the properties of voids found from mock galaxies are compatible with those of voids identified from dark matter haloes. Also, as mentioned above, the use of HOD or semi-analytic galaxy mocks incur further complications in the framework of $f(R)$ gravity, the remedy of which is beyond the scope of the present work. Therefore, as an initial step, }
we only use voids found using dark matter haloes. 

We apply our void-finding algorithm to the halo field in our simulations of GR and 
$f(R)$ gravity. 
We use haloes with the mass cutoffs $M_{\rm min}=10^{12.8} M_{\odot}/h$, $10^{12.868} M_{\odot}/h$, $10^{12.865} M_{\odot}/h$ and 
$10^{12.842} M_{\odot}/h$ for GR, F4, F5 and F6 simulations respectively, where $M_{\rm min}$ is the minimal halo 
mass of the catalogue. $M_{\rm min}$ is chosen to {ensure that the haloes contain at least 100 particles each, and
the slight differences in the $M_{\rm min}$ values for different models is to make sure 
that the halo number densities among different models are the same}. This is necessary because it is likely that different 
populations of voids will be found if the tracer densities are different, even for the same simulations. By using the same number 
densities of tracers for different models, we make sure that the differences among void populations are 
purely due to structure formation of different models. {As mentioned in the previous section, we explore the results in the
halo defined void statistics using different overlap percentages of $50$ to $0$, and study the effects this choice has on
the significance of the comparison between GR and $f(R)$ statistics.}

\begin{figure*}
\begin{center}
\advance\leftskip -0.8cm
\scalebox{0.48}{
\includegraphics[angle=0]{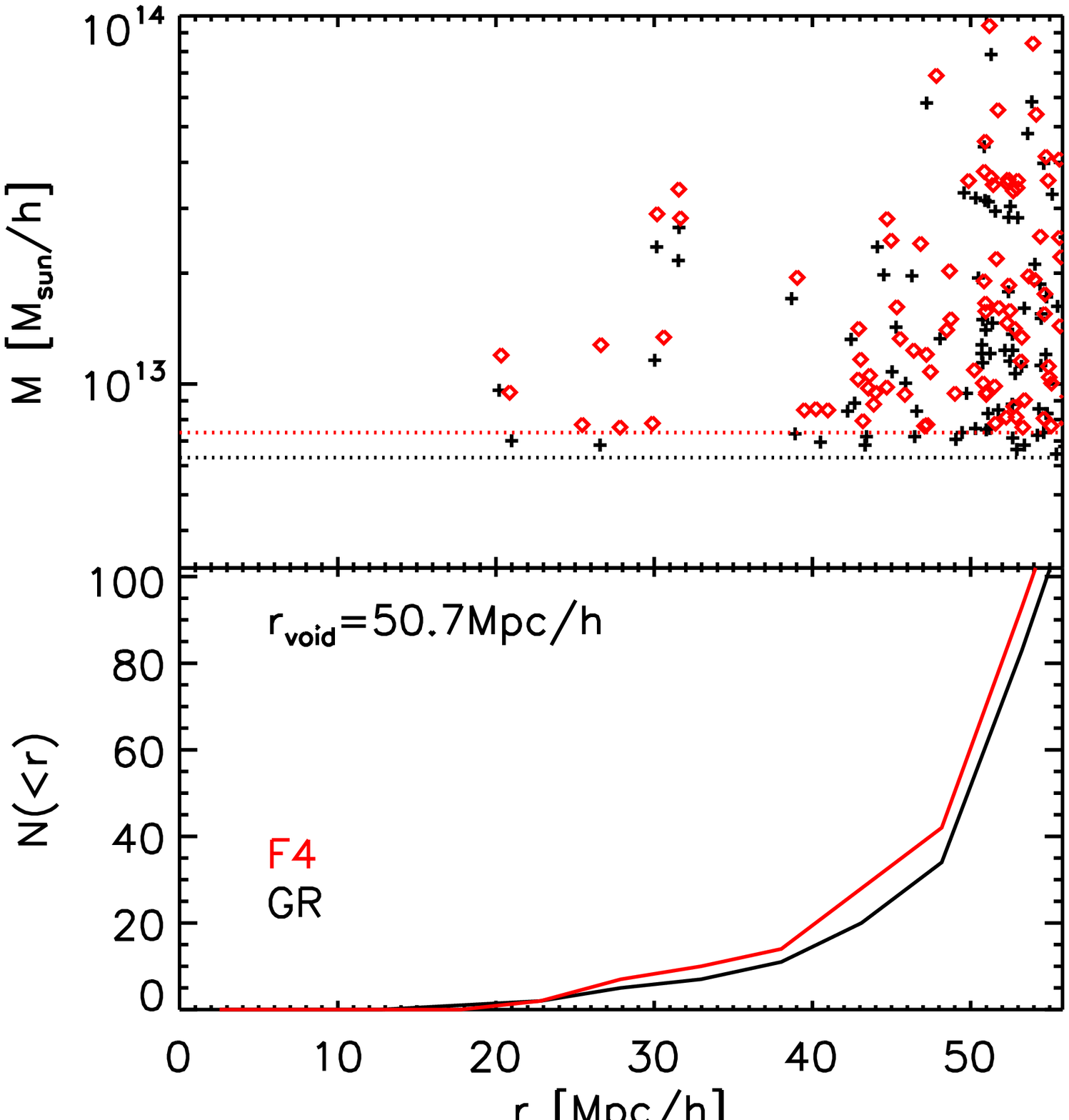}
\includegraphics[angle=0]{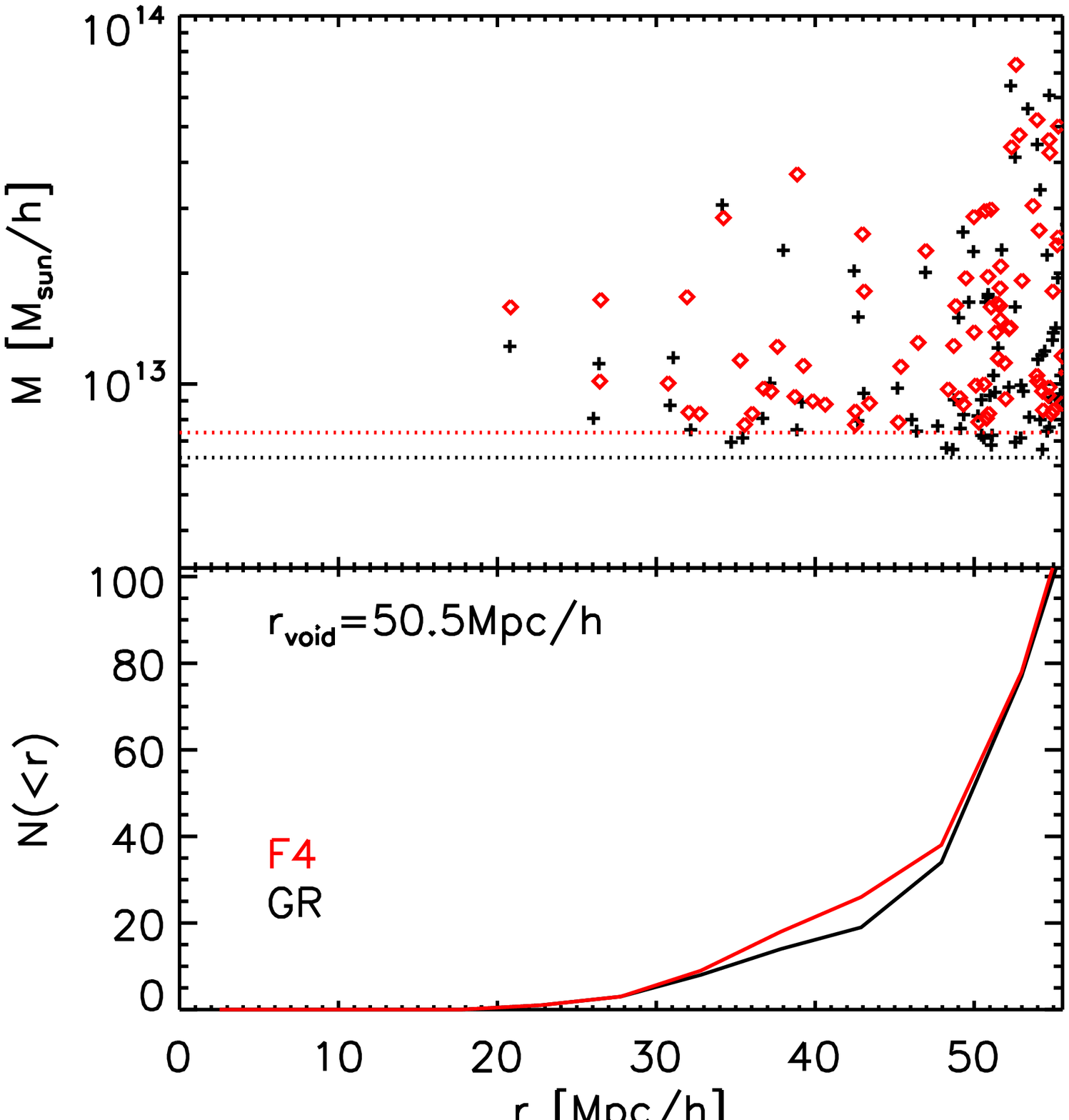}}
\caption{The two largest voids found in our GR simulations using haloes above $10^{12.8}M_{\odot}/h$, 
as indicated by the black-dotted lines. On the top panels, each black cross represent{s} the 
radial distance of a halo from the void centres versus the halo mass. The red diamonds are the 
haloes found within the same comoving radius from the same GR void centres, but from the simulation for the F4 model; the halo mass 
cut for this model is indicated by the red dotted line. The slight differences between the black dotted 
and red dotted lines are to make sure that the number density of haloes between these two models are the same. 
Bottom panels show the cumulative number of haloes from the centre of the GR voids. The F4 curves being above 
the GR curve suggests that more haloes form within these void regions in F4 than in GR.}
\label{fig:LargestVoids}
\end{center}
\end{figure*}

{The resulting void abundances for halo voids {with $20$ percent overlap} are shown in Fig.~\ref{fig:dndrhaloes}. The results 
have several important features. First, since haloes are highly biased tracers of dark matter, and are mainly distributed along walls and 
filaments, voids found in this way are more abundant and larger than those identified from the dark matter field. 
Second, the difference of the void abundances amongst different models is much smaller than in the case 
of dark matter voids, and this suggests that it is much more difficult to use the observed void abundance alone to constrain 
$f(R)$ gravity. {This is possibly a consequence of the fact that haloes are highly biased tracers of the underlying 
dark matter field and have formed from the high peaks of the initial density field: while the fifth force can drastically 
change the clustering of dark matter, its effects on haloes are weaker and a density peak which corresponds to a halo 
in $f(R)$ simulations is also likely to have formed a halo in the GR simulation.} 
Third, and most interestingly,} {we notice that voids larger than} {$r_{\rm void}\simeq25~$Mpc/$h$} {are less abundant in $f(R)$ models than in GR.}

The lower abundance of voids with $r_{\rm void}\geq25~$Mpc/$h$ in $f(R)$ is counter intuitive and may seem inconsistent with the physical picture 
that voids grow larger and emptier in $f(R)$ gravity because the fifth force points towards the edges of 
voids \citep{Clampitt2013}. {To check this, we use the void centres found in the GR simulation and count all the haloes residing 
within the GR void radius, for the different models. Two examples comparing F4 and GR halo number counts in this way 
are presented in Fig.~\ref{fig:LargestVoids}. It is interesting to note that there are more haloes found {in the $f(R)$ simulation} within the {same} 
void radii, even {when} the minimal halo mass cut is lower for GR (as indicated by the dotted lines). This is likely due to 
the stronger environmental dependence of halo formation in $f(R)$ models \citep{LiEfstathiou2012}. The fifth force in $f(R)$ gravity in 
these large under-dense environments {is} stronger, which makes it more likely for haloes of the same mass to 
form in $f(R)$ than in GR. The halo number density within the same sphere {is} therefore smaller in GR than in $f(R)$, making 
it easier to pass our void selection criteria in the former case. We can also see that the
masses of haloes in F4 are larger. Thus even if we make the same $M_{\rm \min}$
cuts for both simulations, the same phenomenon that the largest voids in GR are bigger than 
in $f(R)$ {gravity {would still be present}, as we have checked explicitly}.}

At the largest void radii, $f(R)$ abundances seem to approach the GR values.
Given the considerable error for the void abundance at large $r_{\rm void}$, this may simply be a statistical fluctuation. However, we remark here that such a behaviour might also be physical. Even though the fifth force makes halo formation more efficient 
in $f(R)$ voids, the density in the largest voids could be so low that 
not many new haloes form after all, and the relatively small number added to the halo number density still 
allows these voids to be identified as in GR. This possibility needs to be investigated with 
care in larger volume simulations, which have better statistics.

For small voids, model differences for the abundances are also smaller, as also reflected by the differences of void sizes on the right-hand sub-panel of 
Fig.~\ref{fig:dndrhaloes}. It is possible that small voids are emptier of haloes by definition, since our void finding algorithm ensures that they contain no haloes 
above our mass threshold within $r_{\rm void}$. The abundance of cosmic web structures within small voids is likely to be lower than in large voids. 
The fifth force in small voids expels mass out of the void region, but fails to trigger much more halo formation, simply because there are not many 
structures in it for new haloes to form. Therefore, the differences between $f(R)$ and GR in terms of void abundances may be smaller at small radius. 

Quantitatively, by adopting as error the scatter of the abundance from $1$ Gpc/$h$ sub-volumes of the $1500$ Mpc/$h$ 
simulations, we can estimate the signal-to-noise ratio of the fractional difference between $f(R)$ models and GR (the scatter is 
shown as shaded areas in Fig.~\ref{fig:dndrhaloes}). The S/N peaks at approximated 30 Mpc$/h<r_{\rm void}<40$ Mpc$/h$, where void abundance in F4, 
F5 and F6 is lower than in GR with a S/N = $14, 6$ and $\sim 2$ at their peak, respectively, 
as shown in Fig.~\ref{fig:dndrhaloes_SN}.  The significance of the difference clearly depends on the adopted 
value for the minimum void size. 
In principle, with a larger volume, as the statistics of large voids increases, the peak of S/N may 
shift towards larger radius.

In all cases the samples of voids are well centred as evidenced by the $\sigma_4$ values obtained for the voids.  
The inset in Fig.~\ref{fig:dndrhaloes} shows the distribution of $\sigma_4$ for GR; as can be seen it peaks at $\sigma_4\simeq0.05$, indicating 
an accuracy of better than $5$\% the void radius in its centre. The histogram of $\sigma_4$ has a long but very 
low tail extended to larger values. This suggests that the majority of voids identified are perhaps close to spherical. 
Only a small subset of voids has a large value of $\sigma_4$, which are perhaps very different from spherical or are fake voids.
With the $\sigma_4$ value for each void, we can control the quality of our void sample. We will later use cuts in this parameter to 
select sub-samples of improved statistical value.

\tcr{
In summary,  dark matter void abundances are greater in $f(R)$ than that in GR. In contrast, halo voids show the opposite trend, 
that relatively large voids are less abundant in $f(R)$ gravity than in GR, and their differences are much smaller than in the case of dark matter voids. 
This rings an alarm that voids identified from tracers are very different from those from dark matter. We will see this further on in 
terms of void profiles. Moreover, the decrease of halo void abundances in $f(R)$ model may also provide a unique observable 
to break parameter degeneracies between $f_{R0}$ and $\sigma_8$ in the $\Lambda$CDM model. In $\Lambda$CDM, 
the abundance of large voids is expected to increase with the increase of $\sigma_8$. For the same initial conditions, 
$f(R)$ gravity also shows enhancement of power over a $\Lambda$CDM model at late times, 
which mimics an increase of $\sigma_8$ (though the shape of power spectrum in $f(R)$ model is 
more complicated than a simple increase of amplitudes over a $\Lambda$CDM model, see \citep{Li2013} for more details). 
So, if one observes an effective  $\sigma_8$ that is greater than expected from the concordance 
$\Lambda$CDM model, it is not easy to tell whether the Universe is just $\Lambda$CDM with a higher $\sigma_8$, or if
it is $f(R)$. With the halo void abundance in $f(R)$ model being smaller than that in GR, we expect that this degeneracy can be broken. }

\begin{figure}
\scalebox{0.43}{\includegraphics[angle=0]{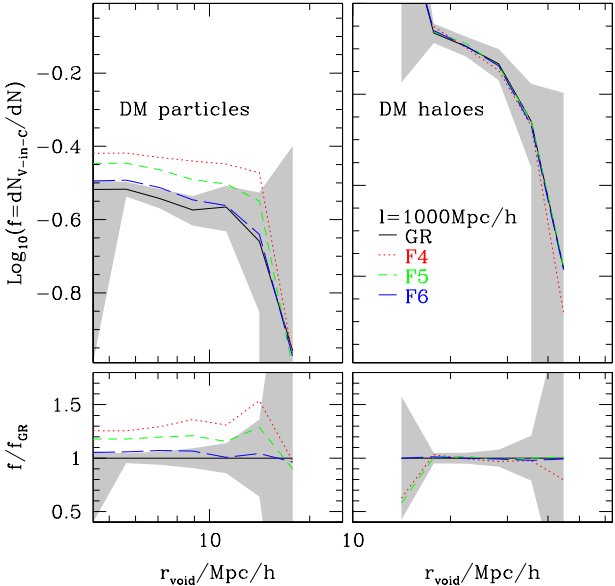}}
\vskip -0.75cm
\caption{Fraction of void-in-cloud voids as a function of void radius, for the GR, F6, F5 and F4 models (different
line types as shown in the key).  The bottom panel shows the ratio of the fractions with respect to GR.}
\label{fig:fvic}
\end{figure}

\subsection{{Void-in-cloud frequency}}

The effect of the fifth force in $f(R)$ gravity makes voids emptier, but this
in turn could result in changes in the frequency of voids that form within over-densities, i.e.
of void-in-cloud voids \citep{Sheth2004}.  With our adopted criterion to separate void-in-clouds
we measure their fraction as a function of void radius, and show the result in Figure \ref{fig:fvic}.

{The left panel shows the results for voids identified using the dark matter particles, and the right panel
corresponds to when using dark matter haloes as tracers of the density field.
As can be seen, the fraction of void-in-clouds in the dark-matter density field
increases from GR through F6 and F5 to F4.  For the latter,
this fraction is $\sim30$ percent higher than for GR, almost constant with void radius.  The differences
are smaller, of only $\sim5$ percent for F6, making this a difficult test on its own to separate
$f(R)$ gravity from GR, but that can provide further constraining power when used in conjunction
with abundances, and also with void profiles, a subject we} {will turn to in the next section}.

{When using dark-matter haloes to identify voids (right panel), the fraction of void-in-clouds is {substantially} higher than for
voids identified using dark matter particles.  However, the differences between
the $f(R)$ models and GR almost disappear, similar to the case of halo void abundances. 

The fact that $f(R)$ gravity produces more void-in-clouds reflects the fact that in $f(R)$ gravity, underdense 
fluctuations in the overdense environments are more likely to develop into voids that are selected by our algorithm, 
because the repulsive fifth force is stronger for void-in-clouds \citep{Clampitt2013}: the consequence of this is more voids are 
generated in overdense regions. Note that the higher void-in-cloud frequency is mainly true for dark matter voids, but not obvious for halo voids:
this is probably due to the sparse sampling of haloes.

\section{Void profiles}
We have seen from the above section that void abundance can be a powerful probe of $f(R)$ gravity when using haloes to identify voids. 
In this section, we shall study the other basic property of voids, their profiles, and see whether stronger constraints can be obtained here. We will look at both the density and velocity profiles of voids.

From this point on we will only use voids identified in the halo fields for two reasons.  The first is to mimic observational data that
consists of galaxies.  The second is related to the findings from the previous section: as results using haloes as tracers produce
smaller differences between $f(R)$ gravity and GR, this choice will provide the minimum possible difference that would need to be detected to tell apart these
different models.

{We start from the halo void catalogue that allows 0-50\% overlapping with neighbouring  voids. The 50\% is the ratio of 
the distance of two void centers versus the sum of their radii. For some extreme cases where one void is much 
smaller than the other, the small ones can still be a sub-void of the larger one, and they both pass our selection criteria. To avoid 
this, by default, we also exclude sub-voids that are 100\% contained by a larger voids. We also exclude voids with shape parameter $\sigma_4>0.2$ to make sure that most of them are close to spherical. In general, changing these criteria does 
not affect most of our results qualitatively, but quantitatively, it may. We will address this in more detail in section~\ref{sec:shear}.}

\begin{figure*}
\advance\leftskip -0.8cm
\scalebox{0.48}{
\includegraphics[angle=0]{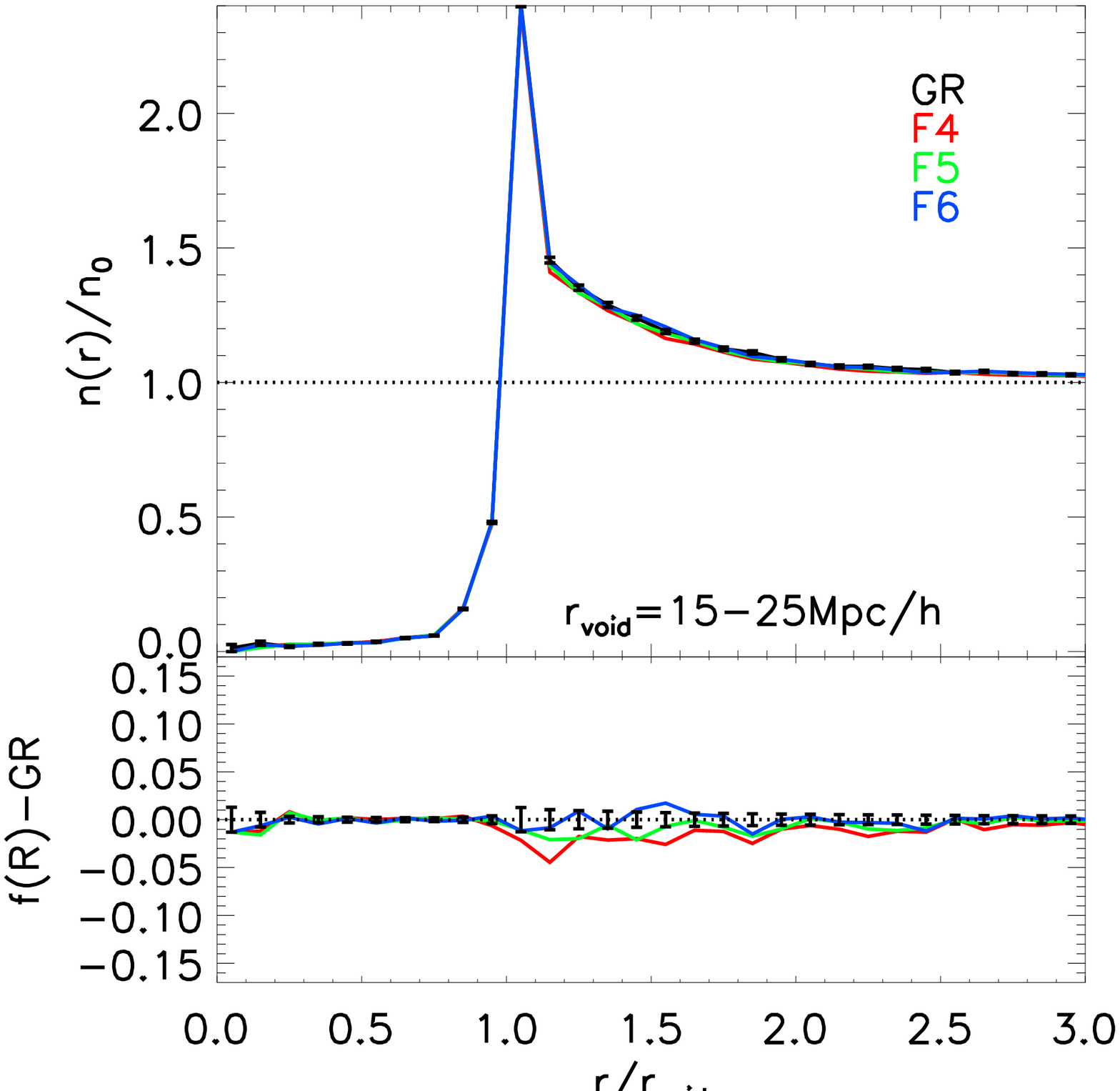}
\includegraphics[angle=0]{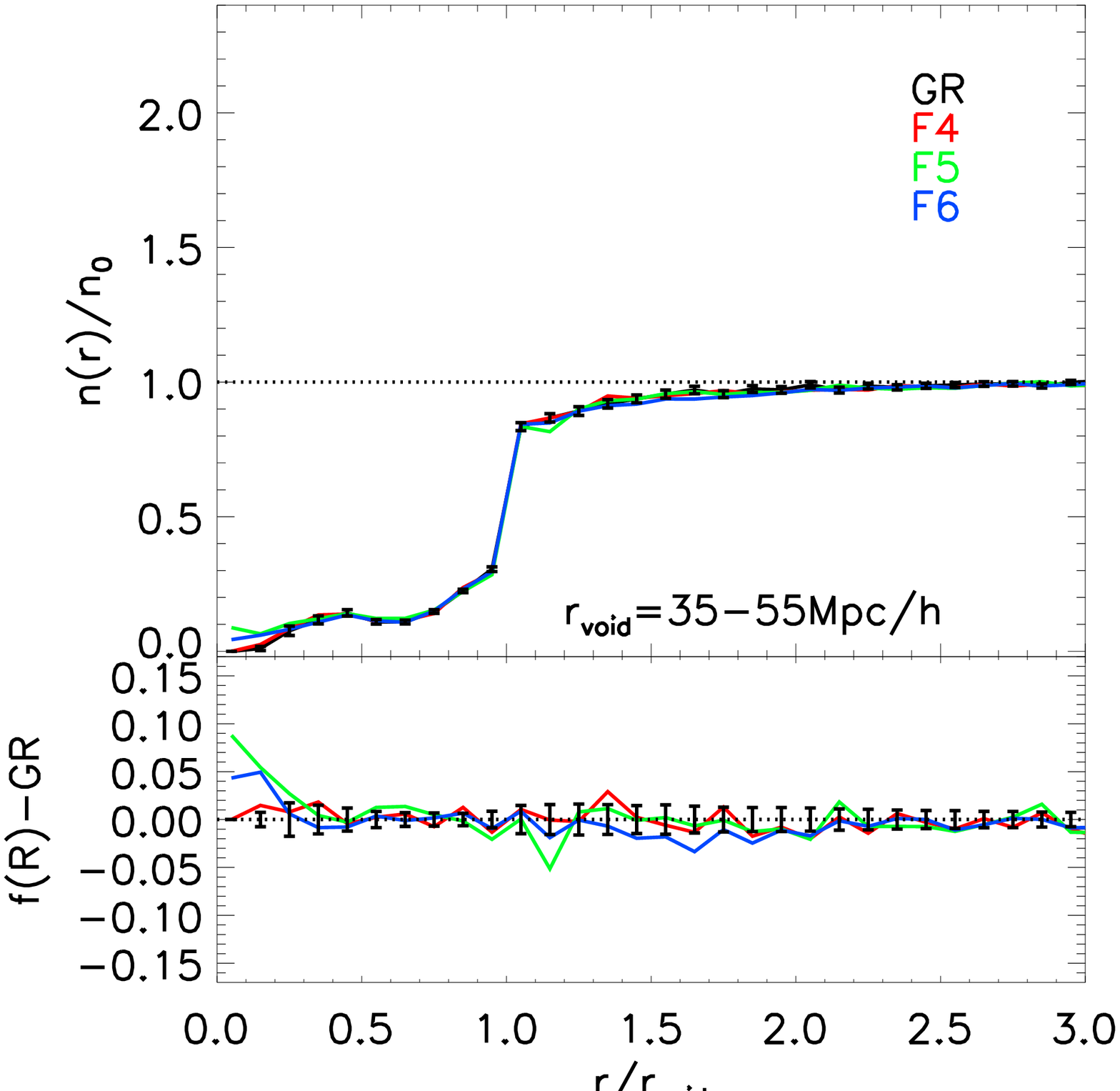}}
\caption{Top panels: void density profiles measured using haloes above the minimum halo mass of $M_{\rm min}\sim 10^{12.8} M_{\odot}/h$ from 
simulations of different models as labelled in the legend. $M_{\rm min}$ is slightly different from $10^{12.8} M_{\odot}/h$ in the $f(R)$ models so that the number of haloes for different models are the same 
(see the text for more details). Error bars shown on the black line (GR) are the scatter about the mean for voids at $15~{\rm Mpc}/h<r_{\rm void}<25~{\rm Mpc}/h$ 
(left) and at $35~{\rm Mpc}/h<r_{\rm void}<55~{\rm Mpc}/h$ (right) found within the 1(Gpc/$h$)$^3$ volume. There are [6038, 5946, 6096, 6307] (left)  
and [296, 323, 319, 261] (right) voids in GR, F6, F5 and F4 models passing the selection criteria. Bottom panels: differences of halo number density profiles of 
voids between $f(R)$ models and GR.}
\label{fig:Haloprofile}
\end{figure*}

\begin{figure*}
\advance\leftskip -0.8cm
\scalebox{0.48}{
\includegraphics[angle=0]{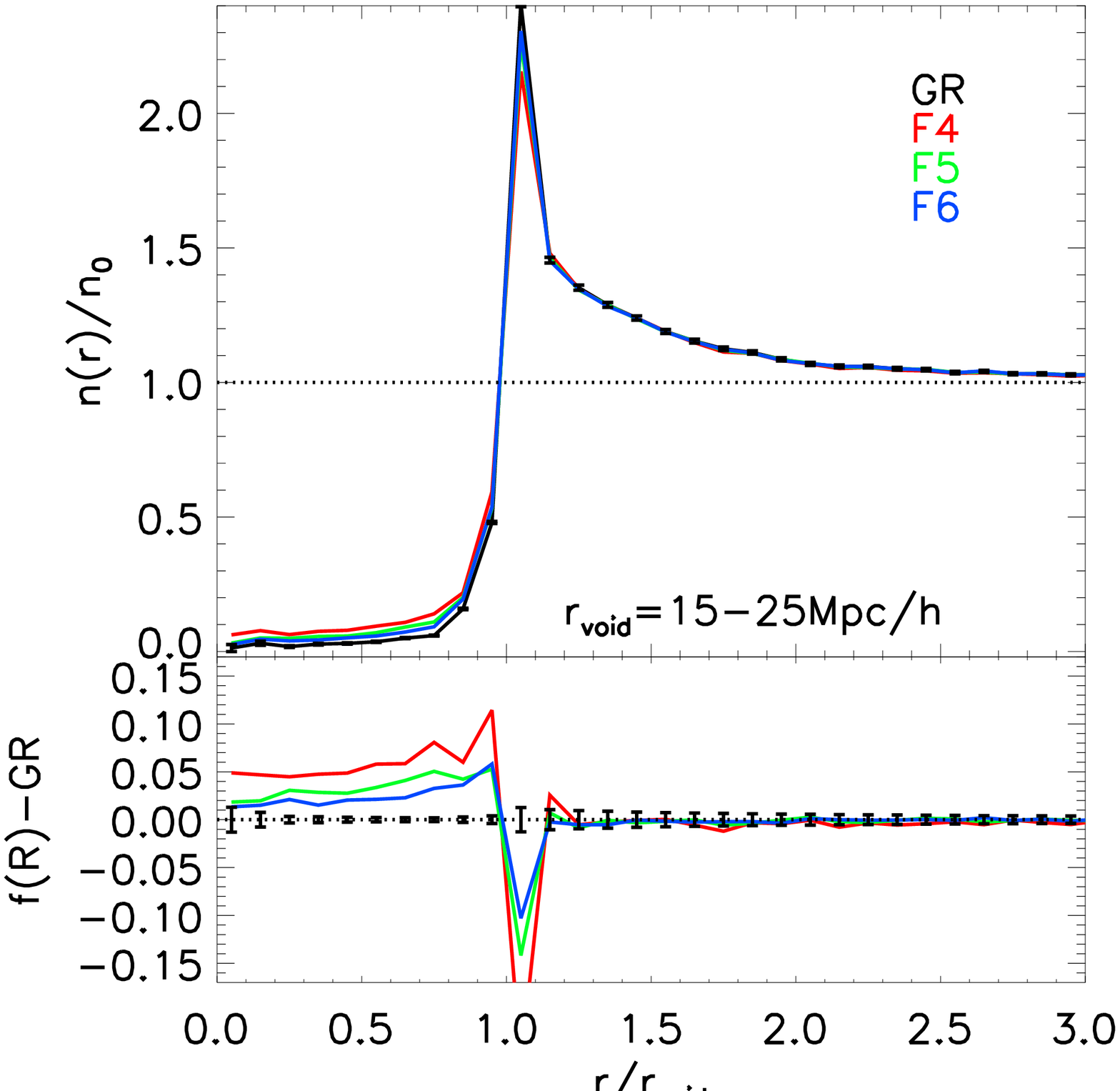}
\includegraphics[angle=0]{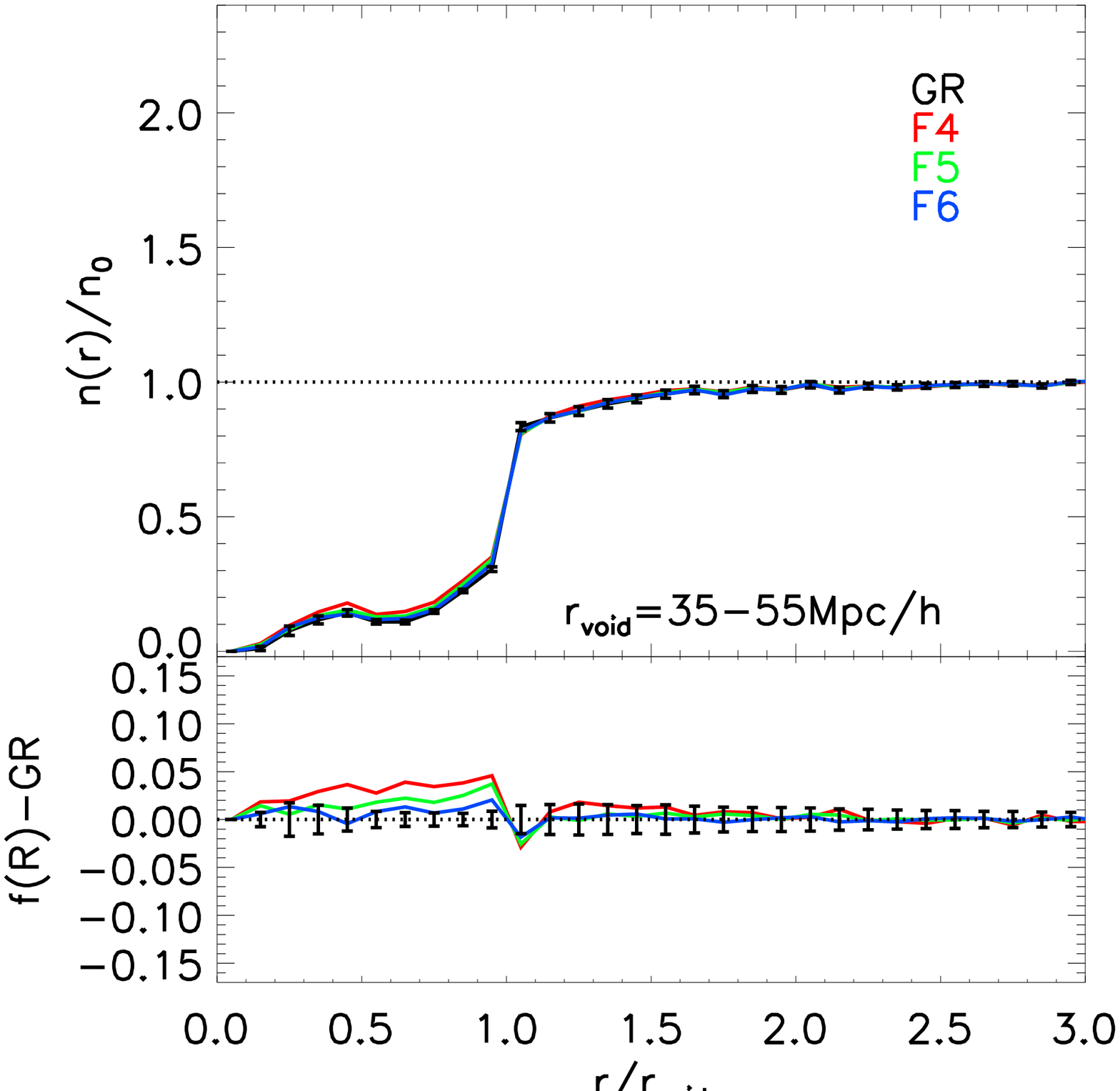}}
\caption{Similar to Fig.~\ref{fig:Haloprofile}, but showing the halo number density profiles using the void centers from the GR simulation and applying them 
to measure the profiles for halo catalogues of $f(R)$ models.}
\label{fig:Haloprofile1}
\end{figure*}

\begin{figure*}
\begin{center}
\advance\leftskip -0.8cm
\scalebox{0.48}{
\includegraphics[angle=0]{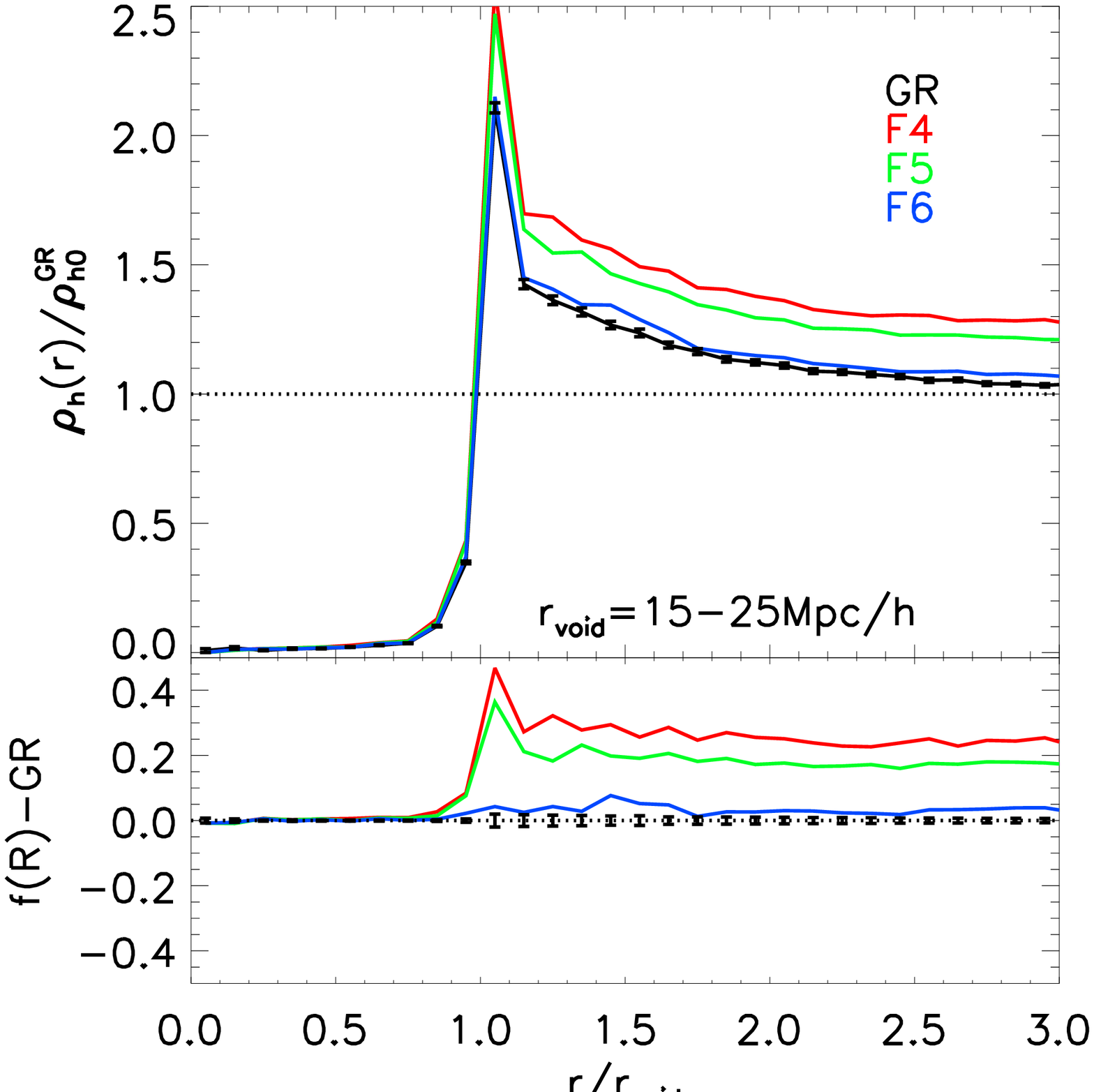}
\includegraphics[angle=0]{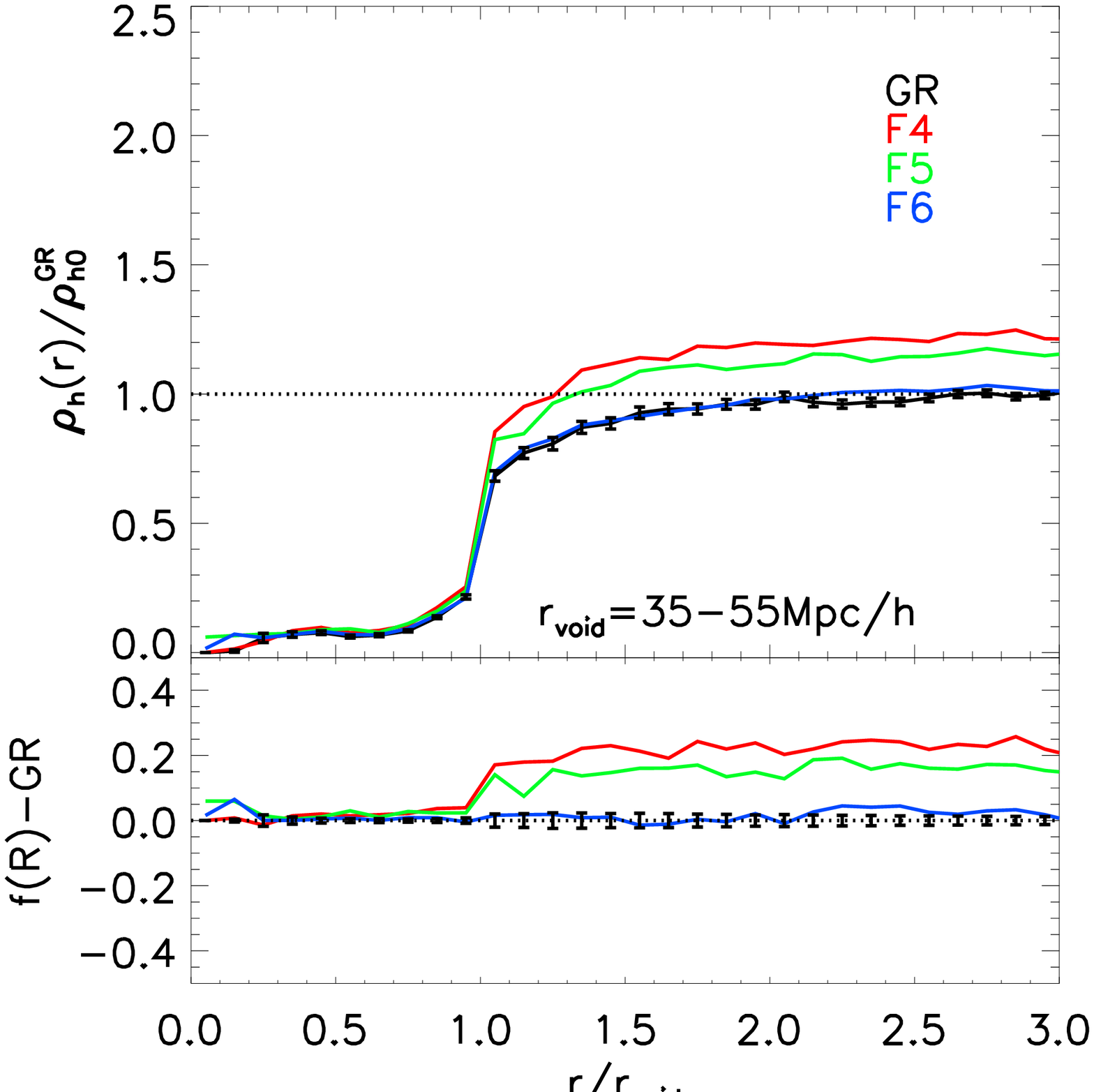}}
\caption{The void profiles for the mass contained in haloes. For better illustration, the profiles are normalized by a common denominator, the mass density of haloes above $M_{\rm min}$ in the GR simulation.}
\label{fig:MassWeight}
\end{center}
\end{figure*}

\begin{figure*}
\begin{center}
\advance\leftskip -0.8cm
\scalebox{0.48}{
\includegraphics[angle=0]{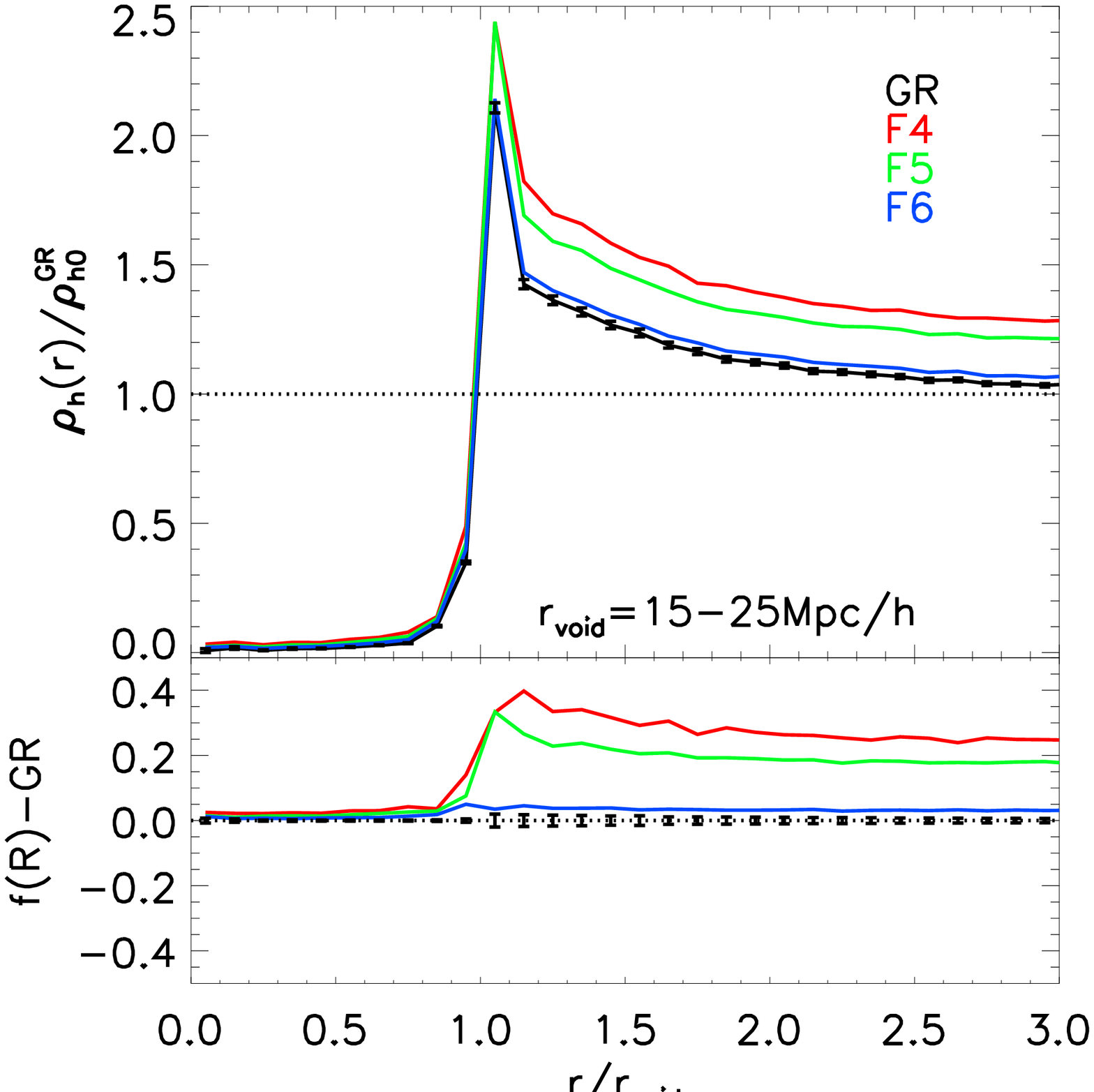}
\includegraphics[angle=0]{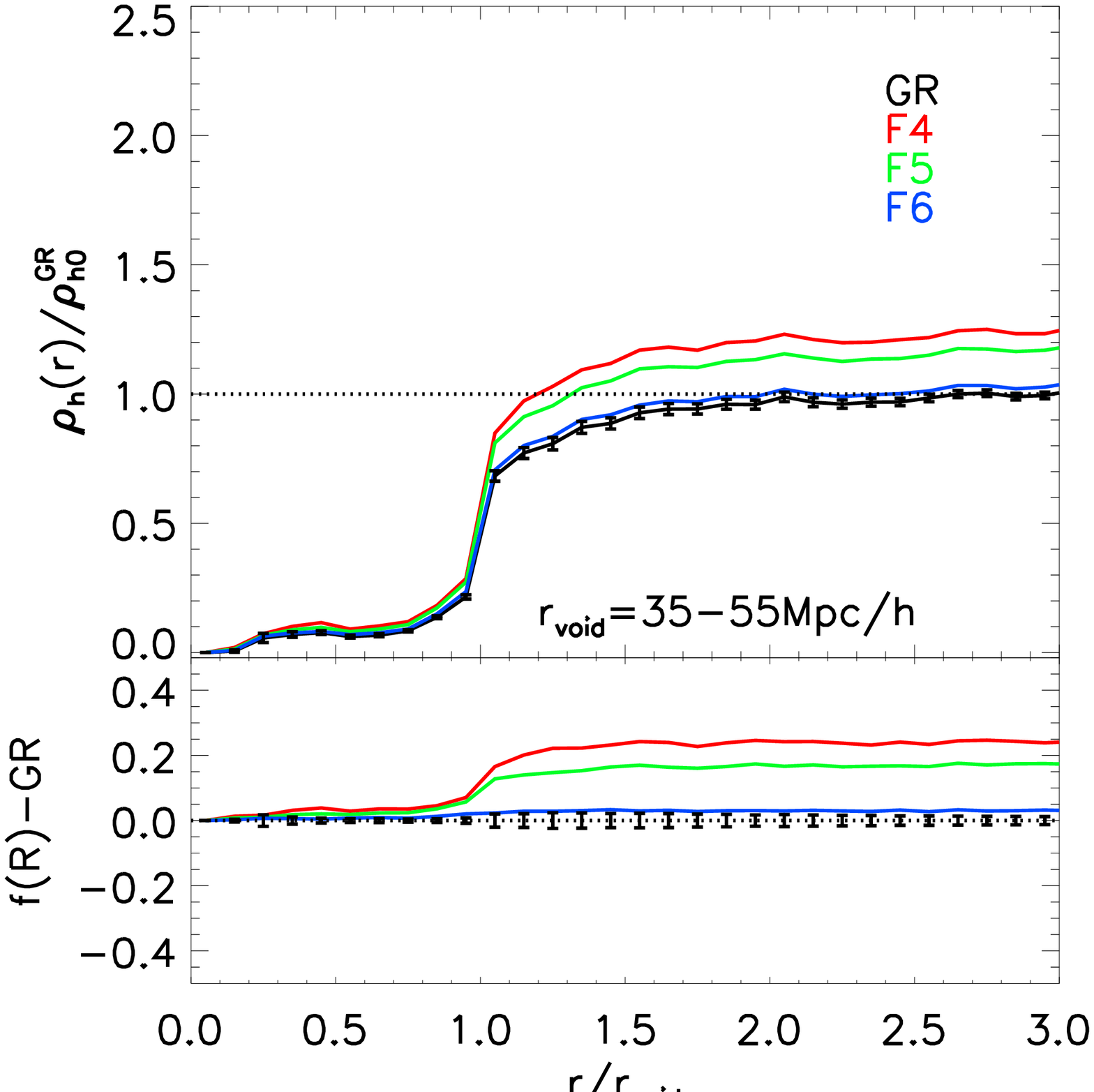}}
\caption{The same as in Fig.~\ref{fig:MassWeight} but showing results from using void centers found in the GR simulation to measure the profiles in all models.}
\label{fig:MassWeight1}
\end{center}
\end{figure*}

\subsection{Void density profiles}
\label{Sec:Dprofile}

 \subsubsection{Halo number density profiles of voids}
 
We first investigate whether there is any difference for the dark matter halo distribution around voids. This has direct 
observational implications as one can simply use the observed tracers (galaxies and galaxy clusters) for this measurement \citep{Padilla2005}. 

With void centres and radii found from halo fields, we count the number of haloes in spherical shells of the thickness of 0.1$r_{\rm void}$ from 
a given void centre and divide them by the comoving volume of each shell to obtain the halo number density profiles of voids. We rescale each {halo-to-void-centre distance 
by the void radius} before stacking them. Results are shown in Fig.~\ref{fig:Haloprofile}, where error bars show the scatter about the mean for the void samples obtained from our 1 (Gpc/$h$)$^{3}$ volume within the radius ranges shown in the legend. Error bars are plotted only for the GR lines for simplicity; the sizes of error bars are similar for the other models, as we have checked explicitly. 
The radius ranges of 15 to 25 Mpc/$h$ and 35 to 55 Mpc/$h$ are chosen to represent approximately the two different types of voids, 
void-in-cloud and void-in-void \citep[see][]{Ceccarelli2013, Cai2013, Hamaus2014}.  

We find that all profiles have very sharp rising features at $r_{\rm void}$. For small voids (left panel), there are striking over-dense ridges at $\sim r_{\rm void}$, and the profiles decrease gradually with $r$ to reach the mean number density at  $r \sim 2.5 r_{\rm void}$. For large voids (right panel), there is no clear over-dense ridge, and the profiles increase asymptotically towards the mean, before reaching it at $r \sim 1.5 r_{\rm void}$. The profiles of the small and large voids are qualitatively as expected for void-in-clouds and void-in-voids \citep{Ceccarelli2013, Cai2013, Hamaus2014}, though there is no clear division in void size for these two types, as we have checked explicitly. In general, smaller voids have steeper profiles than larger ones, with more prominent over-dense ridges. For the very small voids in our catalogues, their over-dense ridges are so prominent that they overcompensate the halo number (or mass) deficit within $r_{\rm void}$. This reflects itself as the inflow of mass towards the void centre beyond a certain radius, as will be shown in Section \ref{Sec:Vprofile}.

It seems that the halo number density profiles of GR and $f(R)$ gravity are extremely close to each other. This is not surprising for the following two reasons. First,  our void finding algorithm requires the number density of haloes to satisfy the same density criteria of 0.2 times the mean, and we do not expect to find strong differences of void profiles at least within $r_{\rm void}$, as shown in the bottom panels of 
Fig.~\ref{fig:Haloprofile}. Second, we are using very massive haloes, which are density peaks, to define voids. Due to the chameleon screening mechanism, the fifth force is likely to be suppressed in over-dense regions, and this minimises the difference of structure formation from GR in over-dense regions. The difference in the halo number density profile, if any, would reflect mainly the difference of environmental dependence for structure formation among different models. 

In general, each dark matter halo does not necessarily correspond to one galaxy. If one uses galaxies as tracers to find voids and measures the resulted galaxy number density profiles of voids, 
the number of galaxies occupied in each halo will put different weights to haloes with different masses. 
This could make the void profiles look different from those presented in Fig.~\ref{fig:Haloprofile}. However, {as shown by \citet{Padilla2005}}, the use of galaxies instead of dark matter haloes has a small impact on the measured void properties. Moreover, assuming that the weights are similar for the different models, the differences between $f(R)$ gravity and GR, which are the main concern of our study, should remain similar. We therefore do not intend to explore beyond using haloes for the present study.

Model differences in void profiles might not be seen if void profiles are self-similar, i.e., if voids in $f(R)$ models grow larger than those in GR without changing their shapes, the rescaled void profiles will look the same. In other words, when looking at voids within the same radius range in different models, one may be comparing different void populations. To check if this happens for the results in Fig.~\ref{fig:Haloprofile}, we test using void centres found in the GR simulation to measure the halo number density profiles for $f(R)$ simulations. This makes sure that we are comparing the same void regions of the same initial conditions. Any differences from GR should be purely due to the different structure formation process in $f(R)$. Results are shown in Fig.~\ref{fig:Haloprofile1}. Interestingly, we find that voids are less empty of haloes in $f(R)$ models than in GR within $r_{\rm void}$. To check if the differences are due to the slight difference in the minimal halo masses among different models, we have also used halo catalogues of the same $M_{\rm min}$ to measure the profiles, and the results shown in Fig.~\ref{fig:Haloprofile1} are confirmed.

Naively, the fact that voids in $f(R)$ models are not as empty of haloes as in GR may seem counter intuitive and contradict the conventional picture of voids in chameleon models. For example, 
the analytical study of \citet{Clampitt2013} 
suggests that voids should expand faster hence become emptier 
in these models. However, it is important to recall that voids defined using tracers (haloes) may be very different from the actual dark matter voids.  When looking at haloes, as we do here, the above conclusion is understandable and consistent with what has been shown in Fig.~\ref{fig:LargestVoids}. Namely, in $f(R)$ models the fifth force is likely to be unscreened in void environments, which makes haloes form earlier and become more massive. Voids therefore become less empty than in GR if they are defined using dark matter haloes of a similar $M_{\rm min}$.

It is also noticeable from Fig.~\ref{fig:Haloprofile1} that the over-dense ridges at $r \sim r_{\rm void}$ in $f(R)$ models  are not as sharp as in the GR case. There are two possible reasons for this. Firstly, since some of these void regions in $f(R)$ models do not correspond to voids in GR, or have slightly different radii from their GR counterparts, including them for the stacking smears the ridge slightly.

 Secondly, for same over-dense regions, especially at the over-dense ridges, the merger rate of haloes may be higher in $f(R)$ models with the help of the fifth force. This makes the number of haloes decrease in those particular regions. If this is true, one may also expect that the mass contained in $f(R)$ haloes above our selection criteria should not be smaller, if not greater than that in GR. To test this, we weigh each halo with its mass to obtain the void profiles of the mass contained in haloes. Results are shown in Fig.~\ref{fig:MassWeight} and Fig.~\ref{fig:MassWeight1}. They are the mass-weighted versions of Fig.~\ref{fig:Haloprofile}  and Fig.~\ref{fig:Haloprofile1}. For better illustration, we have normalised the mass density profiles by a common denominator, which is the density of mass contained in haloes in the GR simulation. Comparing Fig.~\ref{fig:Haloprofile} and Fig.~\ref{fig:MassWeight}, it is clear that although the number density of voids are very close to each other among different models, the masses contained in $f(R)$ models are greater than that in GR. The differences in the mass-weighted profiles are essentially the difference of halo mass functions in these models. Interestingly, even for the case of  Fig.~\ref{fig:Haloprofile1} where 
the number density of haloes in $f(R)$ models are slightly smaller at the over-dense ridge, the mass contained in the haloes are still greater than that in GR (Fig. \ref{fig:MassWeight1}). The difference of the mass-weighted profiles from GR is even greater at the over-dense ridge shown on the left-hand panels.  This supports our argument that in over-dense regions, small haloes merge to make large haloes in a more efficient way in $f(R)$ models.

Note that the numerical experiments shown in Fig.~\ref{fig:Haloprofile1} and Fig. \ref{fig:MassWeight1} are purely for the understanding of the physics in $f(R)$. It is not possible to use voids found in one universe to measure the void profiles of another. What we can observe should be something similar to Fig.~\ref{fig:Haloprofile}, i.e., using tracers of a universe to measure the void profiles in the same universe. In this sense, we conclude that in $f(R)$ models, void profiles seen in halo number density are not distinguishable from the GR results. The situation may be different for dark matter profiles of voids. With this in mind, it is important to understand the connection between void profiles of halo number density and those of dark matter, which is to be addressed in the following subsection.

\begin{figure*}
\begin{center}
\advance\leftskip -0.8cm
\scalebox{0.48}{
\includegraphics[angle=0]{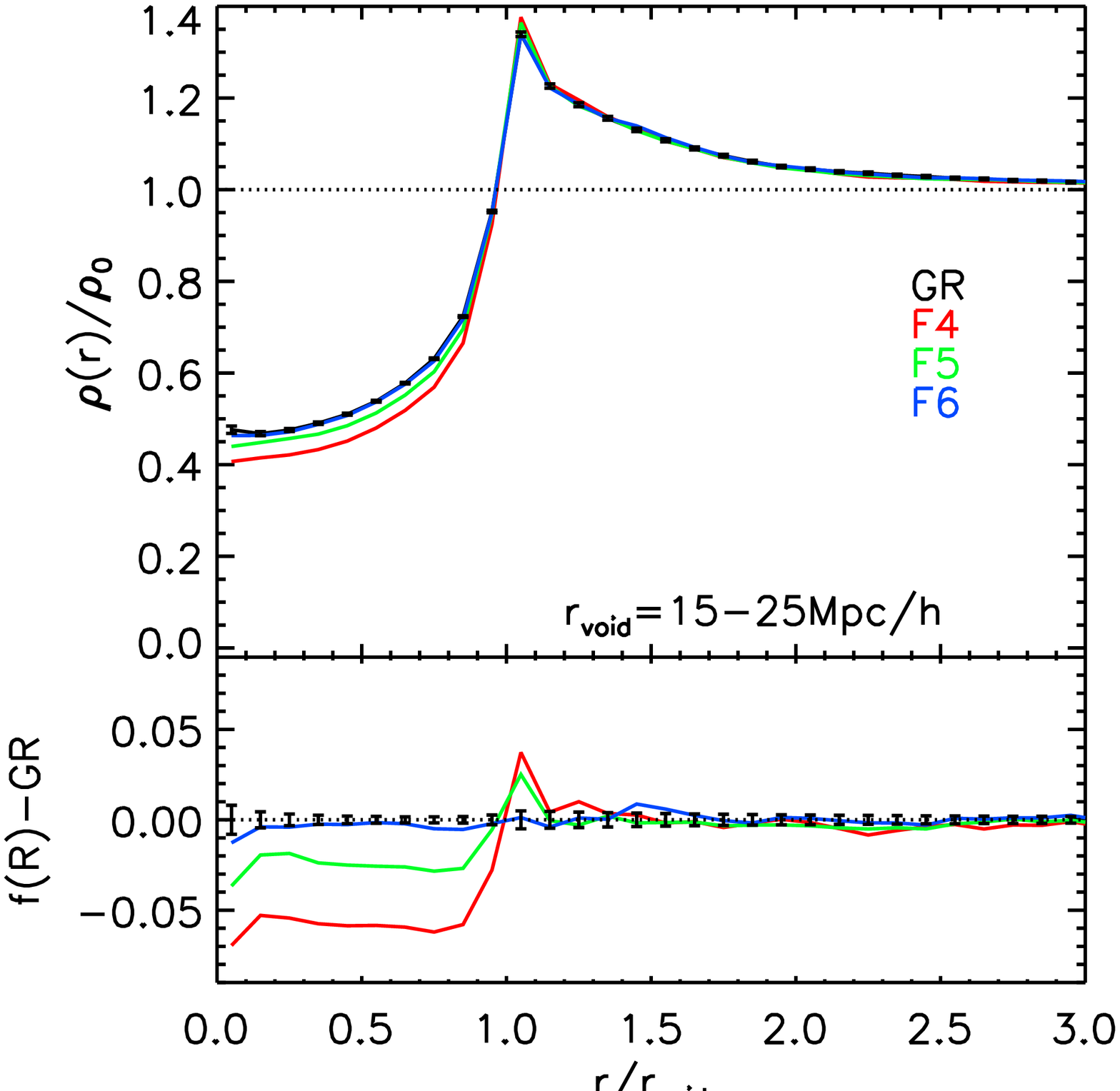}
\includegraphics[angle=0]{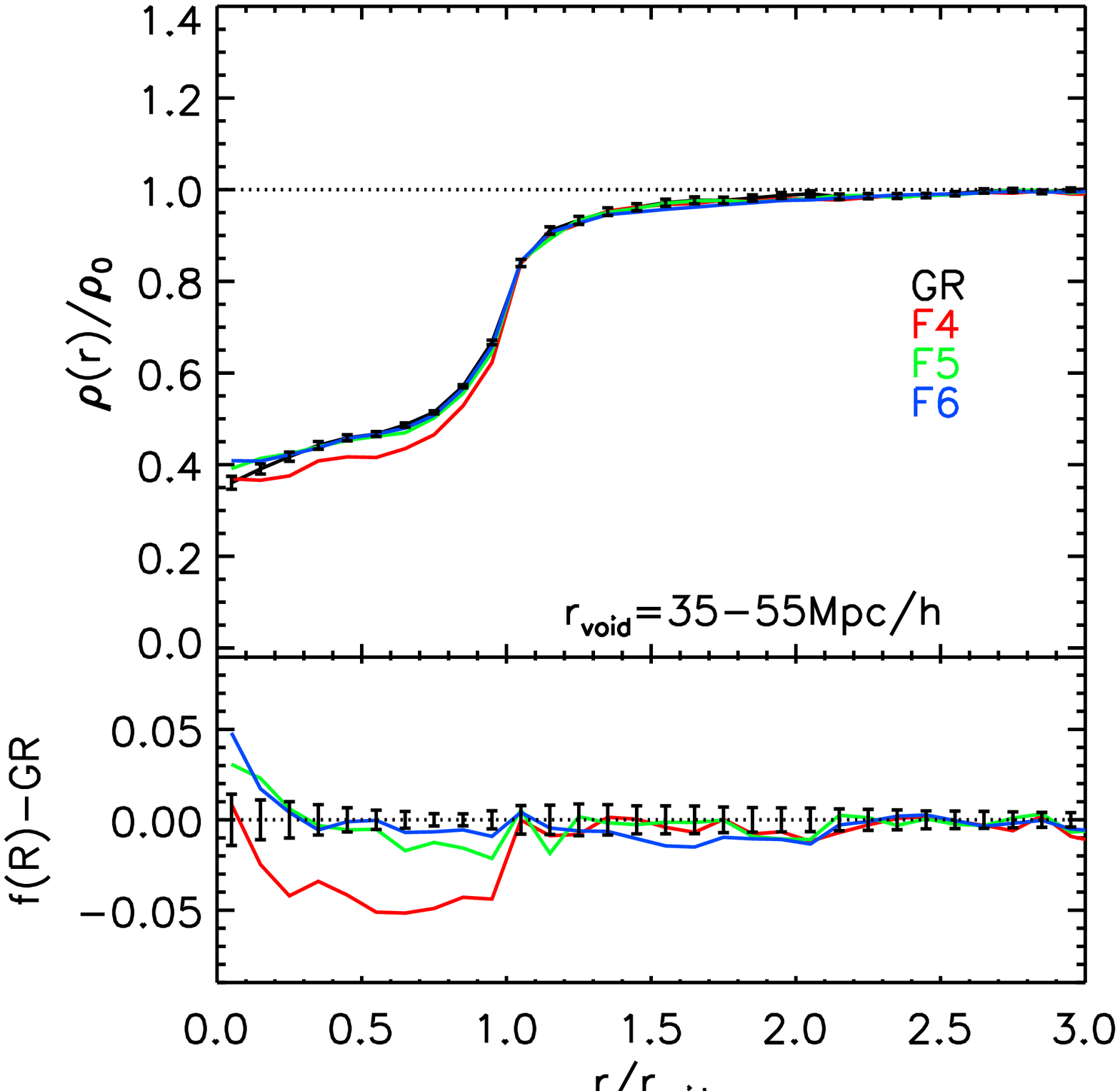}}
\caption{Similar to Fig.~\ref{fig:Haloprofile} but showing void density profiles measured using all dark matter particles from simulations of 
different models as labelled in the legend. Voids are defined using halo number density fields, which are the same as those being used to make 
Fig.~\ref{fig:Haloprofile}.}
\label{fig:Dprofile}
\end{center}
\end{figure*}

\begin{figure*}
\begin{center}
\advance\leftskip -0.8cm
\scalebox{0.48}{
\includegraphics[angle=0]{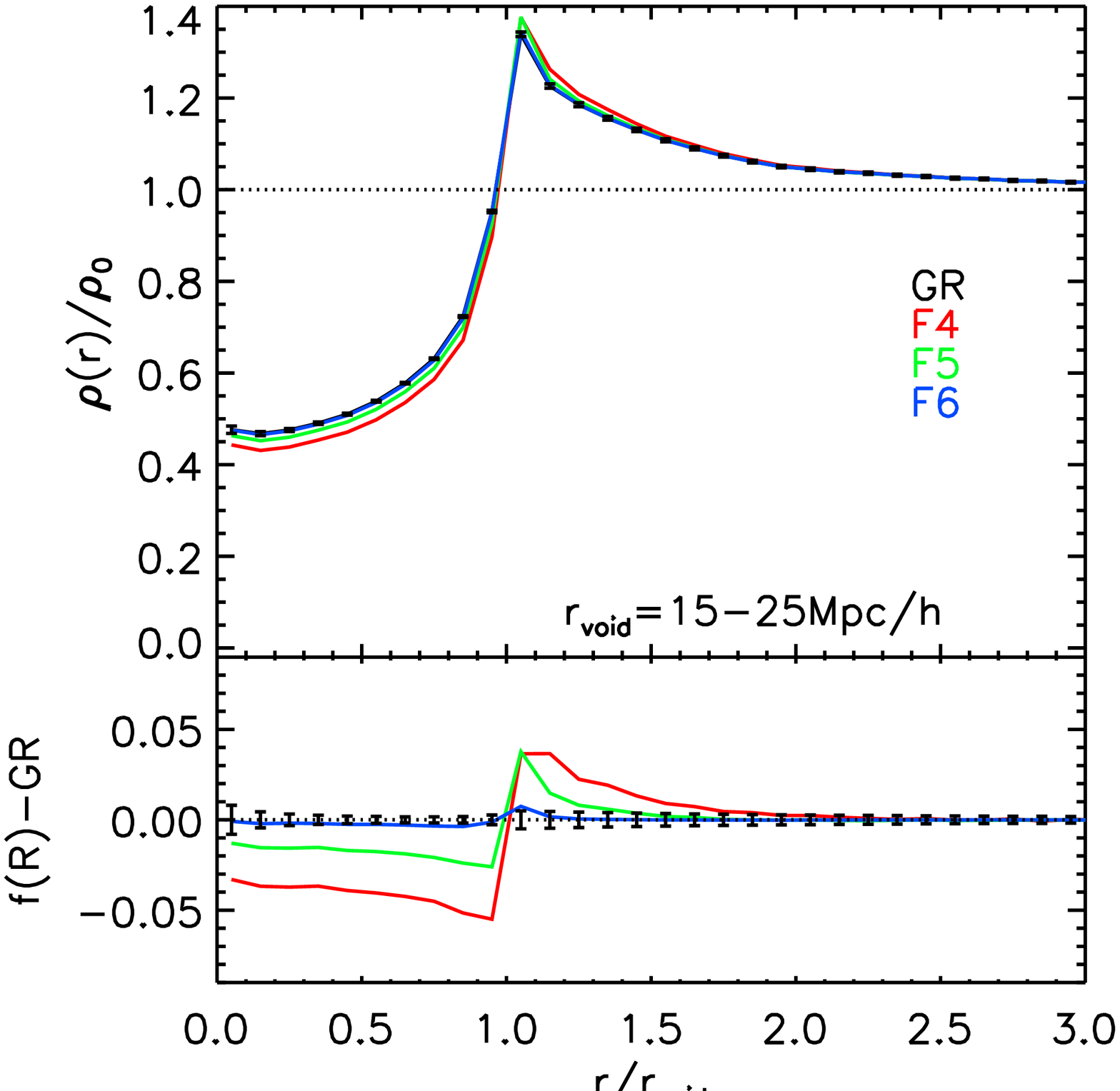}
\includegraphics[angle=0]{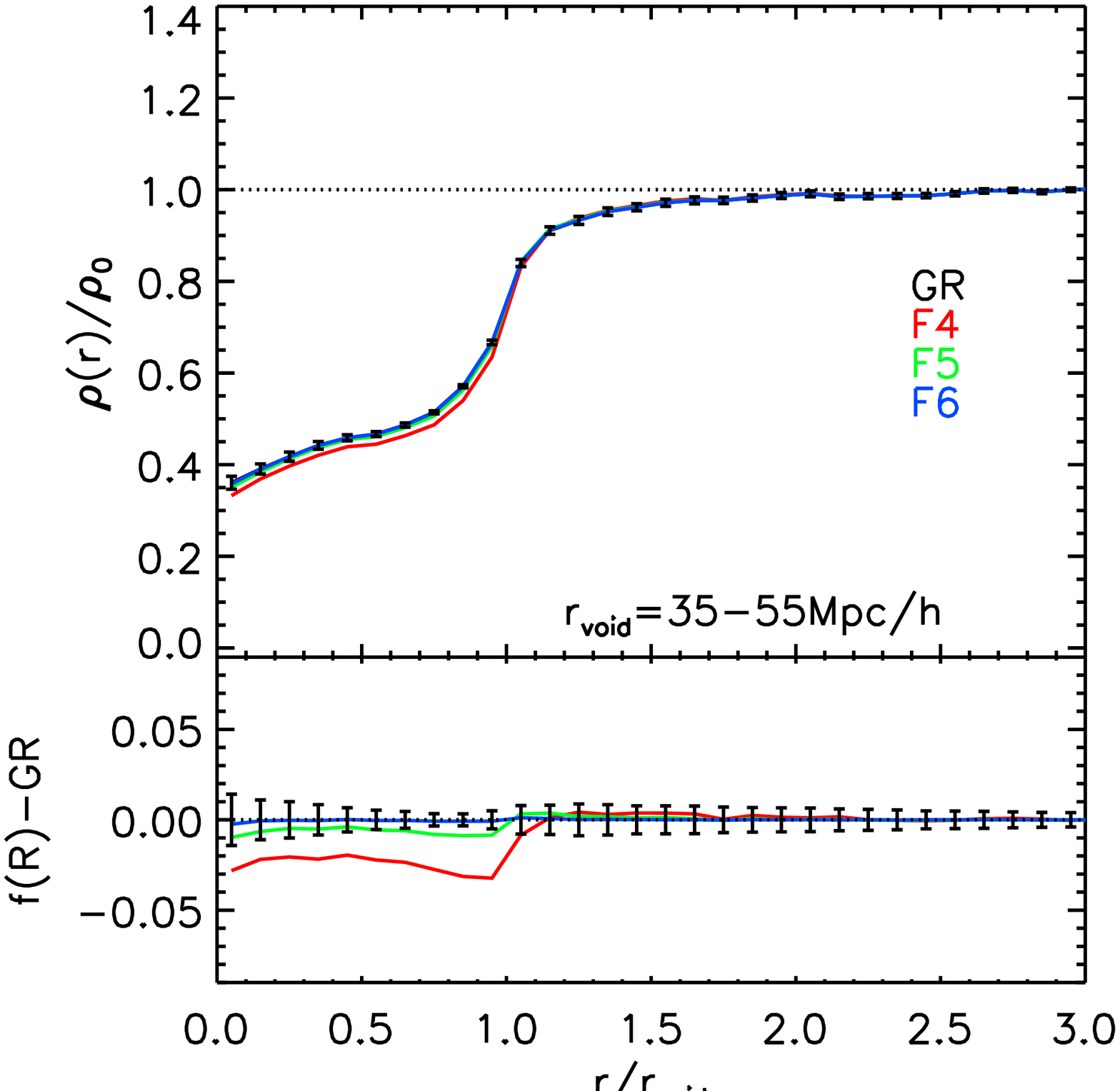}}
\caption{Similar to Fig.~\ref{fig:Dprofile} but showing results using GR void centres to measure void profiles with simulations of different $f(R)$ models.}
\label{fig:Dprofile1}
\end{center}
\end{figure*}

\subsubsection{Dark matter density profiles of voids}

With void centres and radii found from halo fields, we measure the dark matter density profile for voids using all dark matter particles, and rescale them in the same manner 
as for the halo number density profiles. Results are shown in Fig.~\ref{fig:Dprofile}, in which the error bars show the scatter about the mean for the void sample selected from our 1 (Gpc/$h$)$^{3}$ volume within the radius ranges indicated in the legend. Again, we only show error bars for the GR lines for clarity. 

As we have seen above, relatively small voids are much more likely to live in over-dense environments, while big ones are likely 
to reside in under-dense environments. This can be clearly seen in Fig.~\ref{fig:Dprofile} where small voids (left panel) have an over-dense 
ridge at $\sim r_{\rm void}$, as expected for void-in-clouds, but not for large ones (right panel){. These are qualitatively similar to the void halo number density profiles shown in Fig.~\ref{fig:Haloprofile}}. We have found very smooth density profiles with negligibly small error bars for all cases, which reflects the success of our void-finding algorithm in identifying voids using haloes 
that are associated with the underlying dark matter distribution.

For relatively small voids (left panel of Fig.~\ref{fig:Dprofile}), the dark matter density profiles have a very steep rising feature at 
approximately $r_{\rm void}$, and cross the cosmic mean density at about the same place. This is somewhat surprising since the void radius is defined in the halo field by looking 
for the maximal radius that has $n(<r)/\bar{n}\le 0.2$. There is no {\it a priori} requirement in the definition of what the differential or cumulative density profile of dark matter should be like. This can be 
understood as that the halo field traces well the dark matter field in void regions.

More interestingly, it is noticeable that $f(R)$ voids are deeper than GR ones at $r< r_{\rm void}$, and F4 is 
the deepest. At $r> r_{\rm void}$, there is an indication that the over-density ridges are more prominent in $f(R)$ gravity than in GR. This can be more 
clearly seen in Fig.~\ref{fig:Dprofile1}, where void profiles in different models are measured using void centres found in the GR simulation. 
Overall, we can conclude that voids in $f(R)$ gravity are emptier  {(in terms of dark matter)} at $r< r_{\rm void}$, 
even though their halo number densities are the same  {(cf.~Fig.~\ref{fig:Haloprofile}) or follow the opposite trend (cf.~Fig.~\ref{fig:Haloprofile1})} in the same regime. Meanwhile, the over-density ridges may be slightly more prominent in $f(R)$ than in GR, even though it is just the opposite for the halo number density profiles  {(cf.~Fig.~\ref{fig:Haloprofile1}, lower left panel)}. 

The behaviour of dark matter density profiles are now in good agreement with our expectation for $f(R)$ model. Theoretical calculations 
have found that the fifth force in void regions is repulsive in these models, 
driving the expansion of walls of voids to go faster, and making voids emptier than in GR \citep{Clampitt2013}, which is what we see in Figs.~\ref{fig:Dprofile} and \ref{fig:Dprofile1}. 

The situation for relatively large voids (the right-hand panels of Fig.~\ref{fig:Dprofile} and Fig.~\ref{fig:Dprofile1}) is similar to that of 
the small ones except that there is no over-dense ridge for the void profiles. After rising sharply at $r \sim r_{\rm void}$, the profiles gradually 
approach to the cosmic mean, before eventually reaching it at $\sim$1.5 $r_{\rm void}$. The differences among different models for large voids 
are also somewhat smaller, {especially when the void profiles in all models are calculated using GR void centres (Fig.~\ref{fig:Dprofile1}). This agrees with the analytical calculations of \citet{Clampitt2013} that the model differences in terms of void expansion is greater in void-in-clouds than void-in-voids.}

The good agreement between Figs.~\ref{fig:Dprofile} and~\ref{fig:Dprofile1} suggests that the difference we find in Fig.~\ref{fig:Dprofile} is physical.
Nevertheless, the test {of Fig.~\ref{fig:Dprofile1}} is a thought experiment that does not happen in real observations. If we select voids within a certain 
range of radii, their stacked profiles would encode the differences of void populations as well as their {evolution} history. 
In this sense, the results shown in Fig.~\ref{fig:Dprofile} have more practical meaning.

In summary, the results of this subsection unveil the complexity of voids in $f(R)$ gravity. There are clear differences between the profiles of voids identified using tracers and 
those identified using dark matter. This makes it more challenging to use them to place observational constraints. In the next subsection, we will explore the 
possibility of using weak gravitational lensing of galaxies by voids to constrain these models. {Although observationally the direct identification of} voids can only be {done} using tracers (haloes in our case),
the lensing signal generated by voids are associated to their underlying dark matter distribution. Consequently, the combination of the two may provide perspectives of tighter constraints on $f(R)$ gravity.

\subsection{Lensing tangential shear profiles}
\label{sec:shear}

\begin{figure*}
\begin{center}
\advance\leftskip -0.8cm
\scalebox{0.48}{
\includegraphics[angle=0]{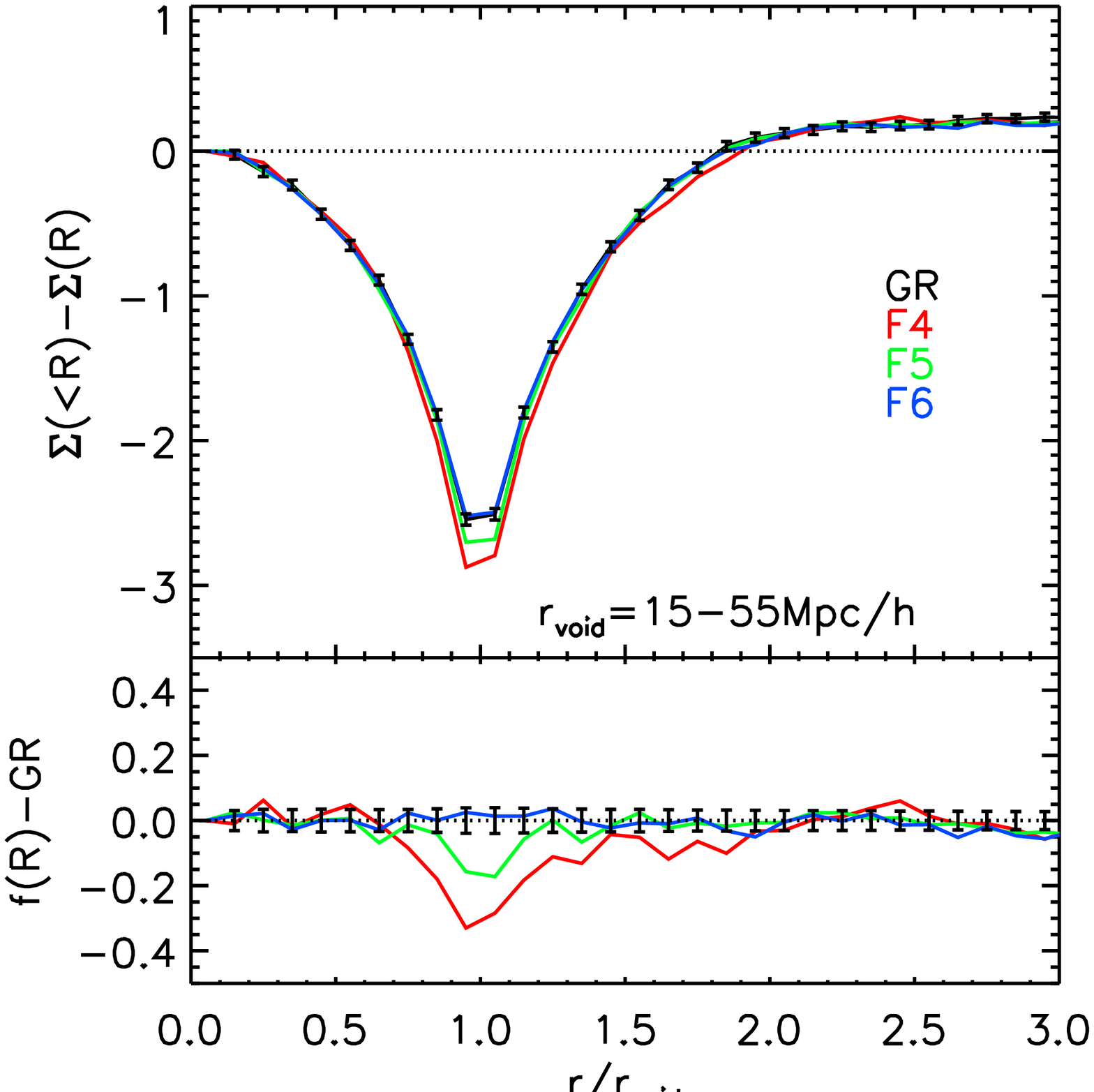}
\includegraphics[angle=0]{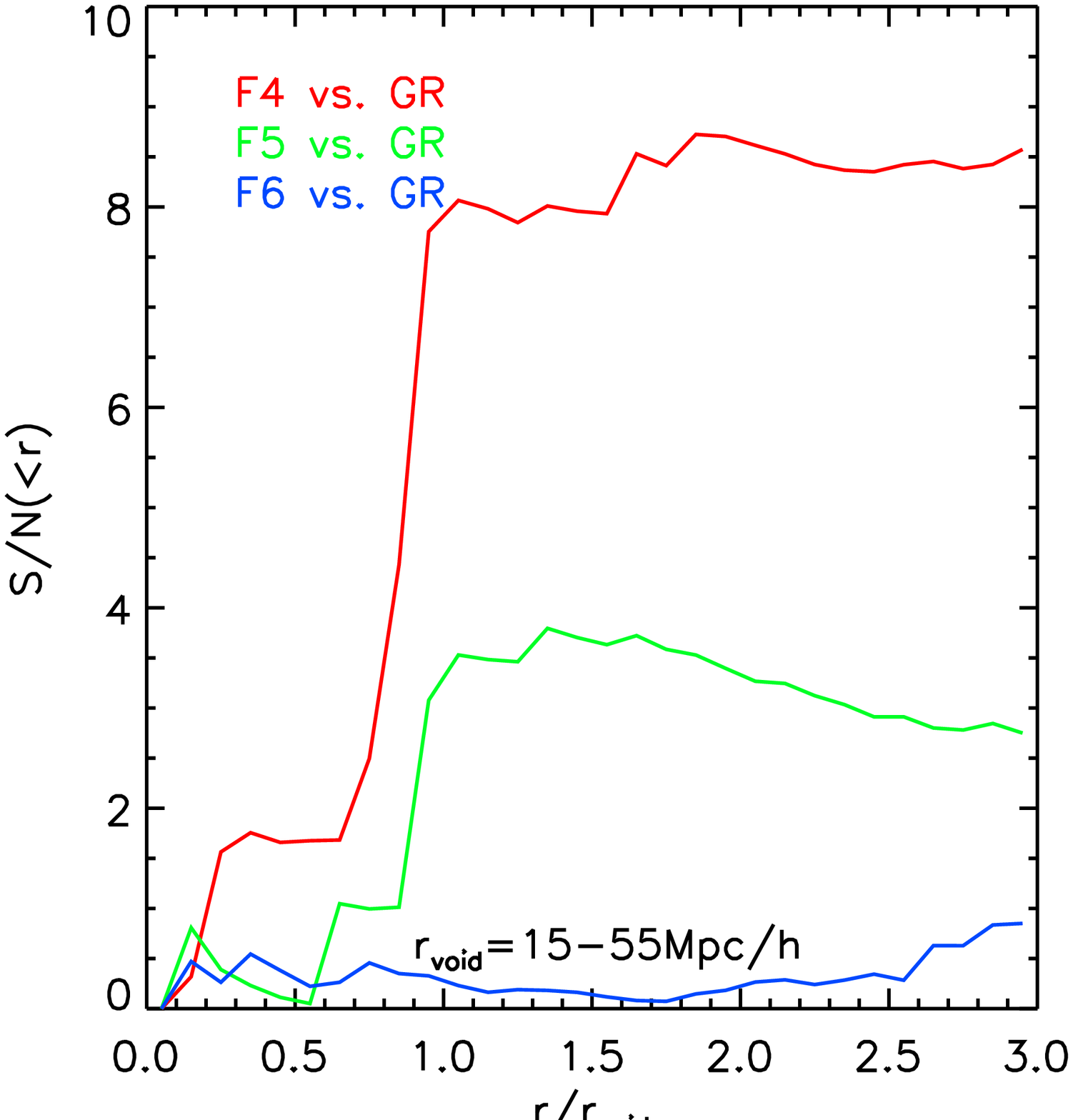}}
\caption{
Left: like Fig. \ref{fig:Dprofile} but showing the lensing tangential shear profiles from stacking all voids with $15<r<55$ Mpc/$h$. 
They are projected over two times the void radius along the line of sight. 
$\Sigma(<R)-\Sigma(R)$ is proportional to the surface mass density within the projected radius of $R$ to which we subtract the surface mass density at $R$. 
Right: the corresponding cumulative (from small to large radius) S/N for the differences between GR and $f(R)$ models.}
\label{fig:Shearprofile}
\end{center}
\end{figure*}

\begin{figure*}
\begin{center}
\advance\leftskip -1.4cm
\scalebox{0.55}{
\includegraphics[angle=0]{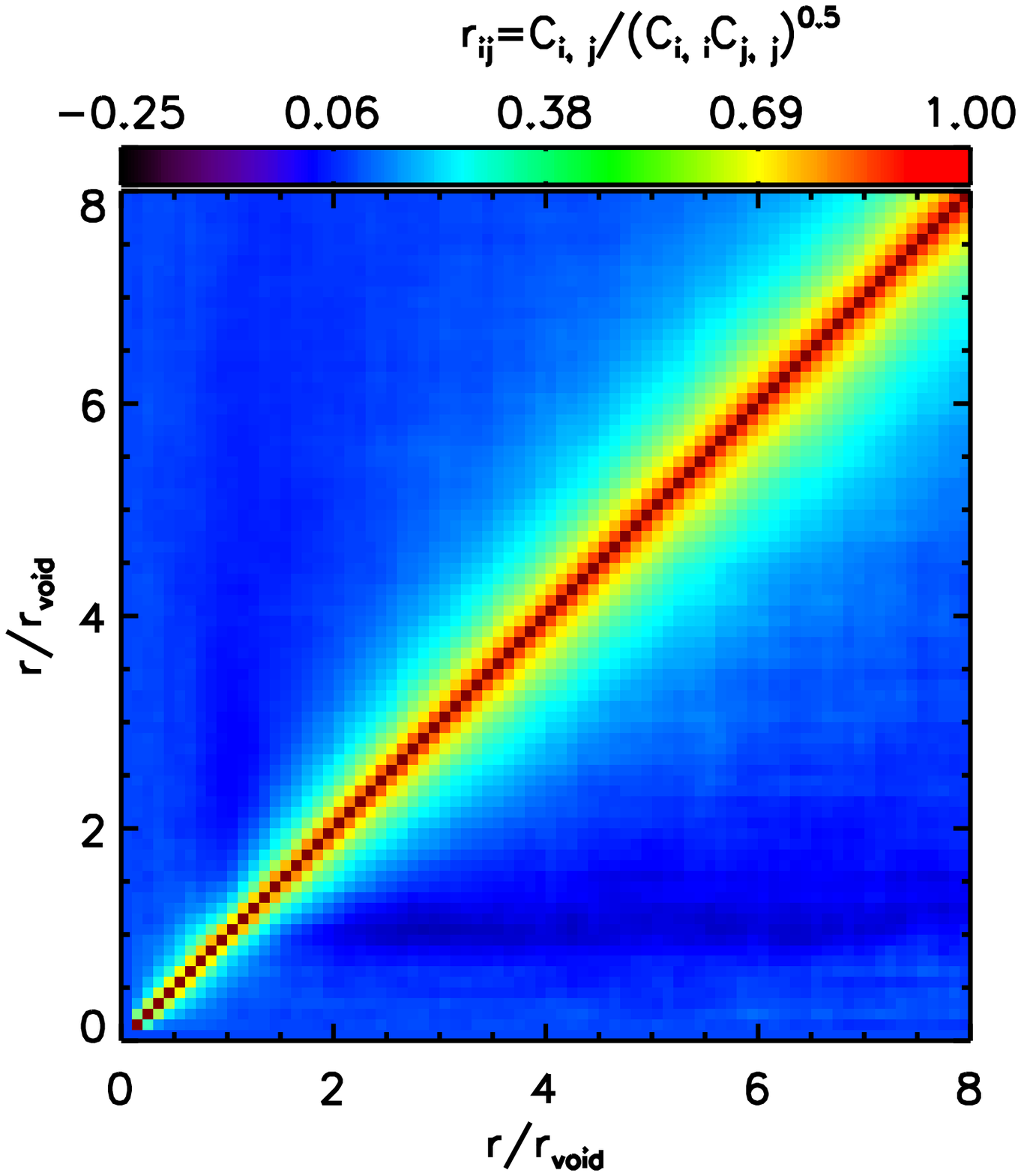}
\includegraphics[angle=0]{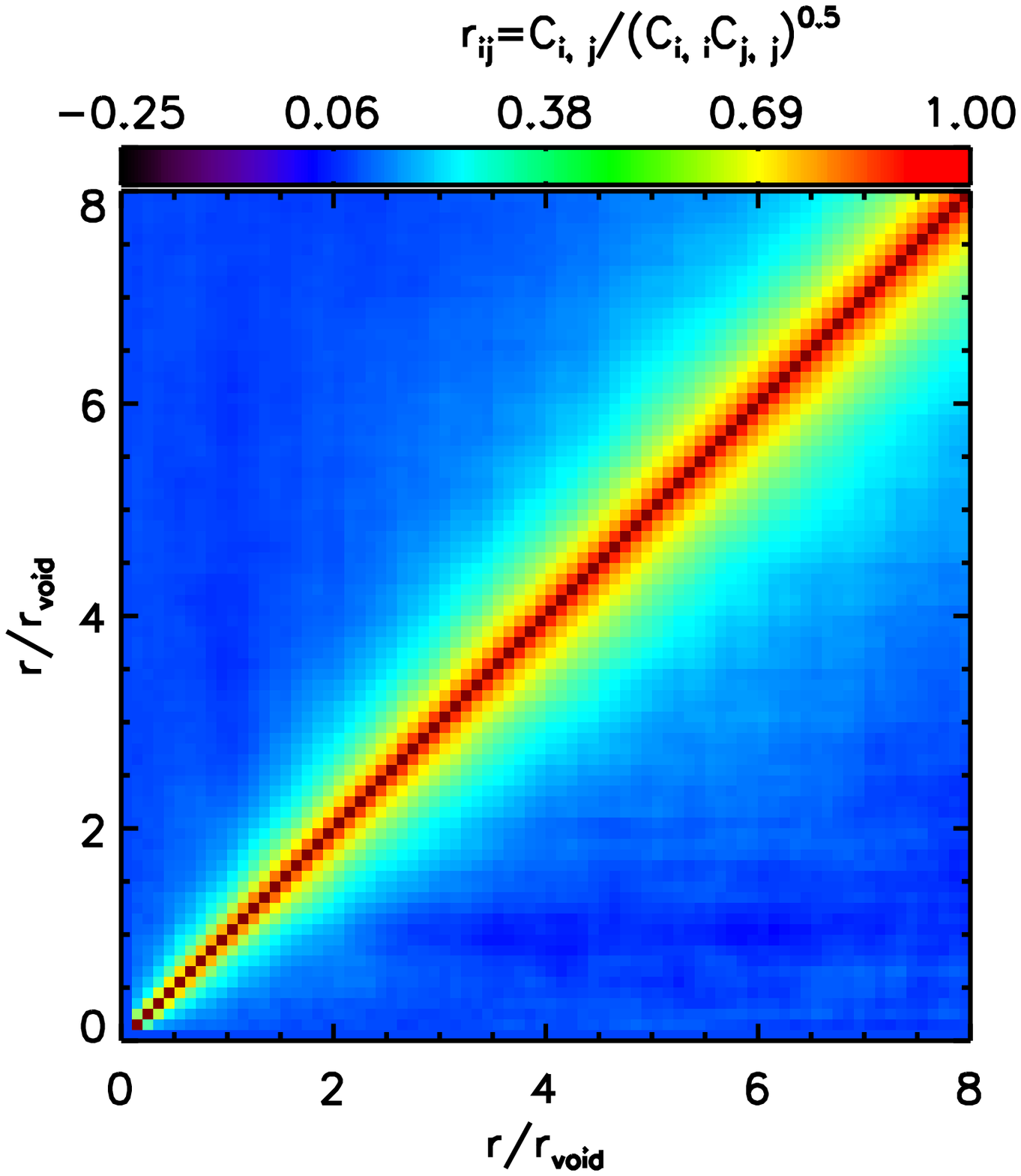}}
\caption{Covariance matrices of the predicted lensing tangential shear from stacking all voids with $15<r<55$ Mpc/$h$ 
from the 1-Gpc/$h$-aside GR simulation. Left: the projection length along the line-of-sight is 2$\times r_{\rm void}$; 
Right: the projection length is 8$\times r_{\rm void}$.}
\label{fig:Covariance}
\end{center}
\end{figure*}

The dark matter profile shown in Fig.~\ref{fig:Dprofile} can in principle be measured using weak gravitational lensing.
Voids defined using tracers (haloes or galaxies) are closely associated with density decrements along the line of sight, as 
mentioned in Section~\ref{Sec:Dprofile}. The deflection of light propagating through voids causes distortions to the shapes of 
background galaxies. The weak lensing shear signal of the background galaxies can be used to measure the matter 
distribution associate{d} with the foreground voids \citep{Amendola1999,Krause2013, Higuchi2013}. 
This technique is similar to using galaxy-galaxy lensing to measure the density profile of dark matter haloes. 
The key quantity that connects observations and theoretical predictions is the tangential shear $\gamma_t$ of the shape of 
the background galaxies. It is proportional to the excess of projected mass density along the line of sight, 
\begin{eqnarray}
\Delta \Sigma(R)=\gamma_t\Sigma_c=\Sigma(<R)-\Sigma(R),
\end{eqnarray}
where $\Sigma_c$ is the geometric factor defined as
\begin{eqnarray}
\Sigma_c\equiv\frac{c}{4\pi G}\frac{D_A(z_s)}{D_A(z_l)D_A(z_l, z_s)(1+z_l)}.
\end{eqnarray}
Here $D_A(z_s)$ and $D_A(z_l)$ are the angular diameter distances from the observer to the source and the lens {respectively,} and $D_A(z_l, z_s)$ 
is the angular diameter distance between the lens and the source. $\Sigma(R)$ and $\Sigma(<R)$ are the projected surface densities
around the centre of a void at the projected distance of $R$ and within $R$. 
They are related to the cross-correlation function of void centres and dark matter via:
\begin{eqnarray}
\label{Eq.ProjectMass}
\Sigma(R)=\bar\rho \int[1+\xi_{vm}(\sigma, \pi)d\pi],
\end{eqnarray}
where $\xi_{vm}(\sigma, \pi)$ is the 2D cross-correlation function of void centres and dark matter. 
$\sigma$ and $\pi$ are the transverse line-of-sight separation and line-of-sight direction. For simplicity, 
we drop the $\bar\rho$ factor and work on dimensionless quantities. 1+$\xi_{vm}(\sigma, \pi)$ is essentially 
the 2D density profile of voids. To increase the signal-to-noise, we do a similar stacking as for the 1D profiles. 
Each void is rescaled by its radius before stacking. We have bins along the $\sigma$ and $\pi$ directions of width of 0.1 $r_{\rm void}$, 
and use Eq.~(\ref{Eq.ProjectMass}) to integrate along the line of sight up to 2 $r_{\rm void}$, where the density comes back very 
closely to the background (see Fig.~\ref{fig:Dprofile}). Results are shown in Fig. \ref{fig:Shearprofile}. 
{For simplicity, we show the profiles from stacking all voids in our 1 (Gpc/$h$)$^3$ simulation box.}
The tangential shear signal peaks at $r\sim r_{\rm void}$ where the {1D} matter density 
profiles are the steepest. This is expected as $\Sigma(<R)-\Sigma(R)$ is sensitive to the slope of the density profile. 
Close to the centre of the void, or further away from the void, the density profiles are relatively flat, hence the tangential 
shear signal is small. But the signal can still be significant at a few times the void radius{, as shown in the left-hand panel of Fig.~\ref{fig:Shearprofile}}.

There is little difference between $f(R)$ gravity and GR for the lensing tangential shear profile {close to void centres}. 
To distinguish between different models, it is more promising to use the shear signal at $0.7 r_{\rm void}\lesssim r \lesssim 1.7 r_{\rm void}$, 
where the tangential shear profiles may differ by 20-30\%. Quantitatively, this is illustrated by the cumulative S/N shown on the right-hand panel of Fig.~\ref{fig:Shearprofile}. 
The contribution for the S/N from $r<r_{\rm void}$ is relatively minor. The S/Ns rise sharply at $r=r_{\rm void}$; 
they peak at about $1.5$ $r_{\rm void}$, reaching 4 and 8 respectively for F5 and F4. With the line-of-sight projection, 
F6 is not distinguishable from GR. 

Note that the error bars given are just for the void sample from a volume of 1 (Gpc/$h$)$^3$.
The current BOSS DR11 CMASS sample has an effective volume of 6.0 (Gpc/$h$)$^3$ \citep{BOSS11,Beutler2013,Sanchez2014}. In principle, the error bars shown in our figures should go down by a factor of 2.4 and the significance level should go up by the same factor if the BOSS DR11 CMASS sample is used, on condition that a deep lensing survey covering the same area of the sky is available. The future EUCLID survey \citep{EUCLID} is expected to have an effective volume of $\sim$20 (Gpc/$h$)$^3$, a factor of 4.4 improvement is expected in this case.

With our simulations, we explore the dependence of S/N for distinguishing $f(R)$ and GR models using the 
tangential shear signal on various aspects. Firstly, we find that including sub-voids is still useful to help increasing the S/N. 
With all sub-voids in our catalogue included, the number of voids increases by approximately 76\%. This helps to increase the 
S/N to 7, and 12 for F5 and F4 respectively, but F6 still can not be told apart from GR, as its S/N 
does not improve. Secondly, including voids with larger values of $\sigma_4$ 
(which means including voids that are potentially very different from spherical) also helps to increase the S/N, 
but very mildly. Finally, the projection of large-scale structure could bring extra noises or even bias the lensing signal 
associated with voids. To test this, we have integrated out to larger line-of-sight distances for the predicted tangential shear signal. 
Overall, we did not find the results to be biased when we integrated for more than twice the void radii, but 
the projection of large-scale structure introduced more noise and slightly larger covariances among the 
errors of different radial bins, as shown in Fig.~\ref{fig:Covariance}. This makes it harder to tell apart the different lines for 
different models, especially for large voids. Increasing the line-of-sight projection from 2 to 6 times of void radii 
deceases the S/N by about 30\%. 

The last point adds challenges for accurate measurement of the tangential shear profiles of voids. 
The density contrasts of voids are not as great as those of haloes. In other words, 
the amplitude of the shear signal associated with voids is relatively small, making it more vulnerable 
to noises such as that introduced by line-of-sight projection. This could set the upper bound for the power of 
constraining $|f_{R0}|$ using the tangential shear signal of voids. We have already seen this from the fact that F6 becomes indistinguishable from GR with the contamination of large-scale structure {\it in a 1 (Gpc/$h$)$^3$ volume}. 

Note that we have not included lensing shape noise for this study, since in general, shape noise may be sub-dominant 
compared to the noise from line-of-sight projection of large-scale structure 
for a DETF Stage IV type of deep imaging survey \citep{Albrecht2006}. Especially, the effect of shape noise is even less 
at relatively large radius of voids \citep{Krause2013, Higuchi2013}, which is the region of our interest to 
distinguish GR from $f(R)$ gravity.  For a lensing survey covering 5000~deg$^2$ of the sky, 
with the number density of source galaxies being $n_{\rm gal} = 12~{\rm arcmin}^{-2}$, the $\rm rms$ of the 
ellipticity distribution of galaxies $\sigma_{\epsilon}=0.3$ and a mean redshift 
of  $\sim1$, the lensing shape noise is clearly sub-dominant for the radius  
shown in their Fig.~2, which is about $0.3\times r_{\rm void}$ \citep{Krause2013}. 
For a survey of similar settings, \citet{Higuchi2013} also find that lensing shape noise has little effect at large scales 
[See Table 4, Table 5 and Fig.~6 of \citet{Higuchi2013} for quantitative comparisons for the lensing S/N 
with or without shape noise.] Therefore, in the regime of $r>0.7r_{\rm void}$ where the shear profiles in 
$f(R)$ gravity differs from GR, it is relatively safe to ignore shape noise for a deep imaging survey. At the very precise level, 
the importance of shape noise varies with the specific design of the lensing survey as well as how it is combined with the (spectroscopic redshift) survey 
necessary to identify voids. For example, a deeper lensing survey with 
good image quality tends to have smaller shape noise per galaxy, hence the contribution to the noise covariance 
from lensing shape noise may be relatively minor. We leave it for future studies to figure out the contribution of 
shape noise for particular surveys. 

The first measurements of the lensing shear signal behind cosmic voids 
has been conducted using the SDSS data \citep{Clampitt2014}. In their analysis, voids are found using LRGs. The 
tangential shear signals are measured for the stacked lensing source galaxies behind the stacked void center. 
A $13\sigma$ shear signal associated with the stacking of those voids has been found, which seems surprisingly 
better than expected from \citet{Krause2013}. 

As a final note, it is possible to combine together the S/N from the lensing shear profiles with the one obtained 
from a comparison of the abundances of voids found using dark matter haloes.  Assuming no correlations between the measurement
of the lensing profiles and the abundances, the S/N between $f(R)$ and GR increases to $18, 11$ and $\sim2$ for F4, F5 and F6, respectively, for a reasonable choice of lower limit in void radius for the abundances, and selection of voids for
the shear profiles.  Again, this estimate is valid for a $1($Gpc$/h)^3$ volume. 

\begin{figure*}
\begin{center}
\advance\leftskip -0.8cm
\scalebox{0.48}{
\includegraphics[angle=0]{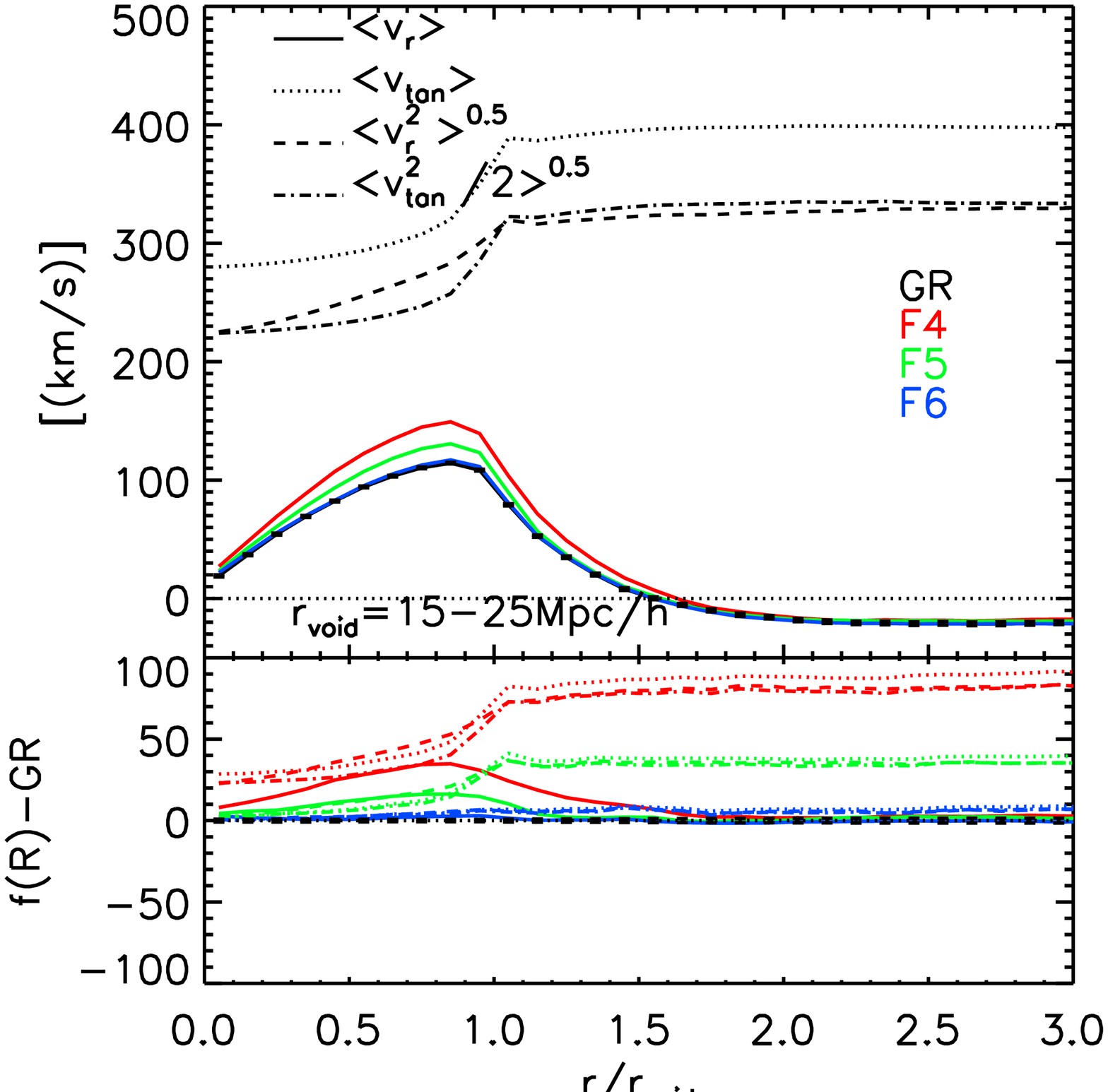}
\includegraphics[angle=0]{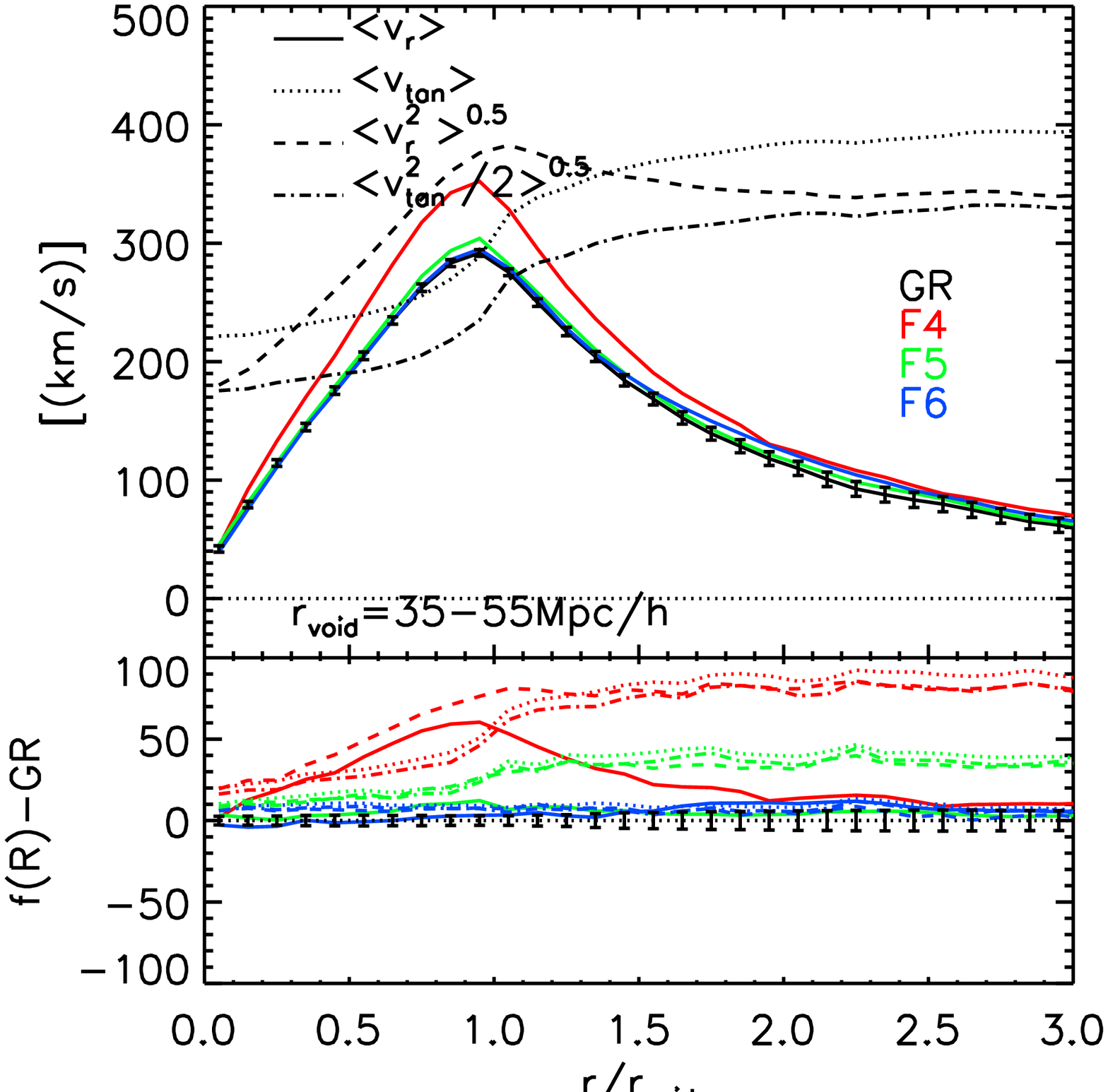}}
\caption{Top panels: different components of dark matter particle velocities  {with} respect to the center of 
voids defined using haloes for simulations of different models labelled in different colours in the legend. 
Solid lines -- the mean radial velocity profiles; dotted lines -- the mean tangential velocity profile;  
dashed lines -- the dispersion of the radial velocities; dash-dotted lines -- the dispersion of half of the tangential velocities. For simplicity, for velocity 
profiles of $f(R)$ models only the radial velocity profiles are plotted. One can appreciate their differences with respect to GR in the bottom panels. 
Error bars shown on the black line (GR) are the scatter about the mean for voids  {of} $15~{\rm Mpc}/h<r_{\rm void}<25~{\rm Mpc}/h$ (left) and  
$35~{\rm Mpc}/h<r_{\rm void}<55~{\rm Mpc}/h$ (right) found within the 1 (Gpc/$h$)$^3$ volume. }
\label{fig:Vprofile}
\end{center}
\end{figure*}

\begin{figure*}
\advance\leftskip -0.8cm
\scalebox{0.48}{
\includegraphics[angle=0]{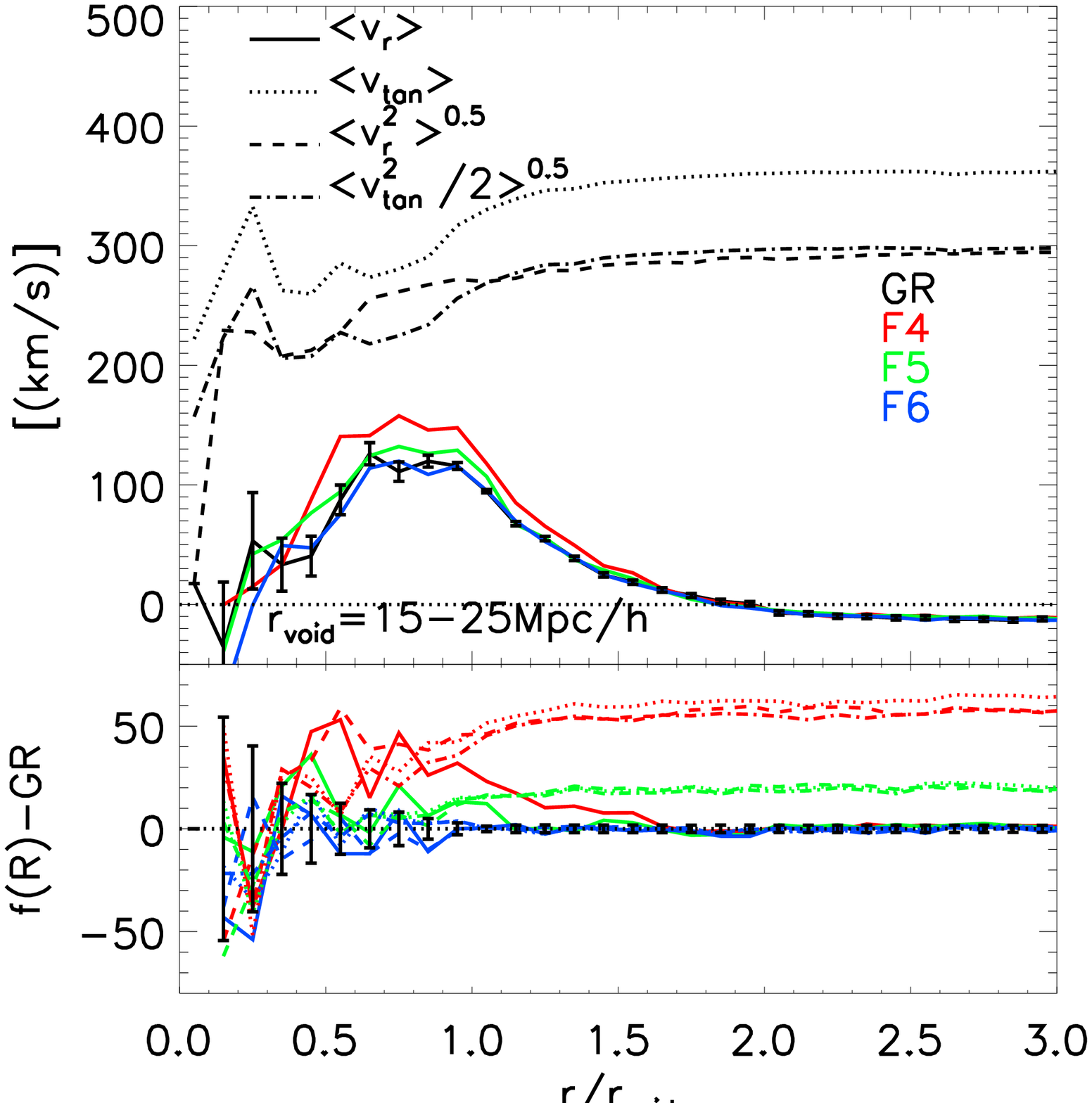}
\includegraphics[angle=0]{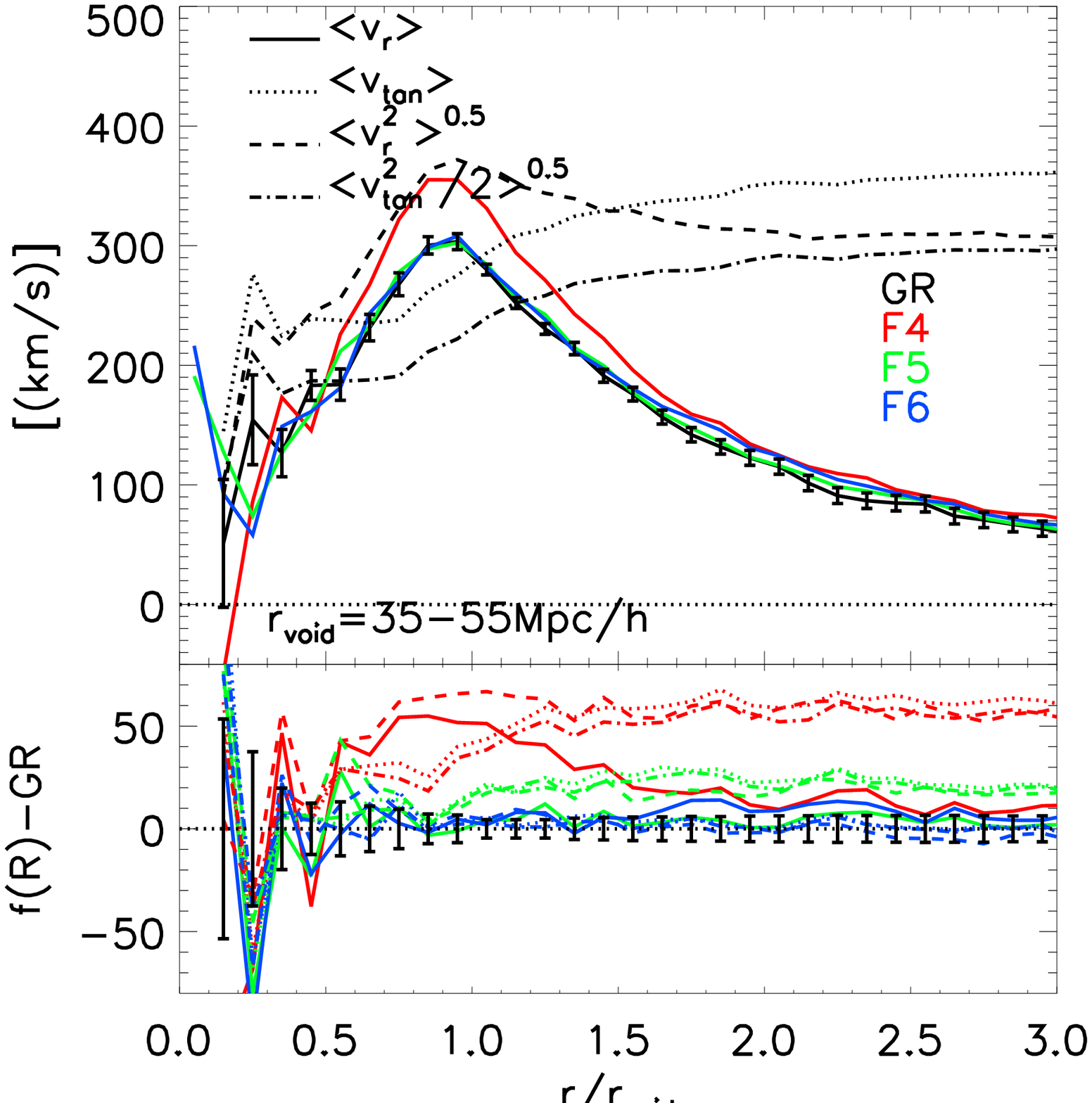}}
\caption{ Similar to \ref{fig:Vprofile} but showing different components of velocity profiles traced by haloes.}
\label{fig:HaloVprofile}
\end{figure*}

\subsection{Void velocity profiles}
\label{Sec:Vprofile}

Velocities of dark matter, as the first integral of forces, should in principle react more sensitively 
to the differences among different gravity models than the density field. To investigate this, we present different components of velocity profiles 
for dark matter in Fig.~\ref{fig:Vprofile} and for haloes in Fig.~\ref{fig:HaloVprofile}. 

The radial velocity profiles, shown as {solid} curves, peak at {$r\sim r_{\rm void}$}, consistent with the 
sharp rise in the density profiles. The sign of the radial velocity being positive means there is coherent outflow of 
mass. For relatively small voids ($r_{\rm void}=15-25$~Mpc/$h$), the outflow is about 5, 15 
and 35 km/s or $\sim$5\%, 15\% and 35\% greater than GR for F6, F5 and F4 at $r\sim r_{\rm void}$. 
Results for larger voids ($r_{\rm void}=35-55$~Mpc/$h$) are shown on the right: they are qualitatively similar to those of smaller voids, 
except that the matter outflows are much stronger. They peak at $r\sim r_{\rm void}$ at about 300 km/s 
for GR, and for F4 it can be greater by nearly 60 km/s. 
Indeed, this level of difference is mildly greater than that in the density profiles shown in 
Fig.~\ref{fig:Dprofile}. The fact that the outflow velocity in $f(R)$ gravity is greater than in 
GR is also expected from the analytical work of \citet{Clampitt2013}: in $f(R)$ gravity, 
the outward pointing fifth force from void centres drives dark matter to evacuate faster from the voids.    

{At $r>r_{\rm void}$, t}he radial velocity profiles decrease.
For small voids shown on the left, they turn negative at $r\sim1.5 r_{\rm void}$, 
which means there is net inflow of mass. This is a clear signature of void-in-cloud configurations. For large voids shown in the right-hand panel, the radial velocity profiles stay above zero, as expected for void-in-void, and the outflow of mass remains beyond $r\sim 3 r_{\rm void}$.

Fig.~\ref{fig:Vprofile} also presents the mean tangential velocity as well as the radial and tangential velocity dispersions. 
It is interesting to see that all these components of velocities have similar levels of differences between $f(R)$ models and GR.  
Fig.~\ref{fig:HaloVprofile} presents the same {results} for haloes. All the above results {regarding}
velocities {in the dark matter field} are confirmed {using haloes}, {although the latter is noisier. We have also computed the velocity profiles for voids identified using GR void centres, and the results are shown in Fig.~\ref{fig:Vprofile1} in the Appendix; there is very little difference from Fig.~\ref{fig:Vprofile}, where the $f(R)$ and GR void centres are independent.}

The velocity dispersion in $f(R)$ models being greater than in GR provides another 
observable for testing $f(R)$ gravity. Stacking of galaxies around central galaxies in 
phase space has been proposed to measure the halo mass at a few times the {virial} radius, and has been applied by \citet{Lam2012} for testing gravity. The level of differences we have found for 
the velocity dispersion between $f(R)$ gravity and GR are $\sim5\%$, $10\%$ and $20\%$ for F6, F5 and F4. These are consistent with 
what was found in \citet{Lam2012} for the case of stacked haloes in phase space for the same models \citep[see][for a test of modified gravity along the same line]{Hellwing2014}.

Before leaving this section, we make the following remarks: 
\newline\indent (i) The differences in velocity profiles are still present when using haloes as tracers, in contrast to the results from the two previous subsections, supporting
the use of tracers to distinguish $f(R)$ gravity from GR. However, model differences in the tangential velocity or radial velocity dispersion are slightly smaller for haloes than for dark matter -- which could be a result of the suppression of the fifth force inside haloes -- indicating the existence of halo velocity bias in $f(R)$ gravity.
\newline\indent (ii) Differences between models in the velocity and density field can be best captured  {by the} the clustering of voids in redshift space. We will conduct detailed studies of voids in redshift space in a separate paper.

\section{Conclusion and discussion}

{To briefly summarise, in this paper we} have studied void properties in $f(R)$ gravity using N-body 
simulations of $f(R)$ \citep{hs2007} and GR $\Lambda$CDM models of the same initial 
conditions and the same expansion history. In $f(R)$ models, the repulsive fifth force in voids drives them to grow larger 
and expand faster. This leads to a {range} of observables that are potentially powerful to distinguish {$f(R)$ 
gravity from GR. In particular, we have found that:}

\indent$\bullet$ {The void abundances in $f(R)$ gravity can differ significantly from in GR, if voids are identified using the dark matter field. More explicitly, the Hu-Sawicki $f(R)$ model with $f_{R0}=-10^{-4}$ (F4) shows a $\sim100\%-400\%$ enhancement of the void abundance over GR in the void radius range of $10\sim20$~Mpc/$h$; the enhancement in $f(R)$ models with $f_{R0}=-10^{-5}$ (F5) and $f_{R0}=-10^{-6}$ (F6) is smaller, but still at $\sim70\%-100\%$ and $\sim20\%$, respectively, in the same radius range. However, if we (more realistically) identify voids using the dark matter {haloes with the same space density}, the difference from GR becomes much smaller, {nearly disappearing at $r_{\rm void}\lesssim 25$ Mpc$/h$, for all three variants of $f(R)$ gravity}. 
By using the expected scatter of void abundances, we find that $f(R)$ can in principle be told apart from GR with a $S/N=14, 6$ and $\sim 2$ for F4, F5 and F6
respectively. Although, in real observations, these may be degraded by the uncertainty of halo mass estimation and the complexity of survey window functions.

\indent$\bullet$ 
We find the counter-intuitive result that, if voids are identified in the halo field, then $f(R)$ gravity produces fewer large voids ($r_{\rm void}\geq30$~Mpc/$h$) than in GR. This is because, thanks to the fifth force, haloes are more likely to form in under-dense regions in the $f(R)$ model, which makes low density regions in this model less empty of halos even though they are emptier of dark matter. \tcr{For the same void regions, $f(R)$ models are less empty of haloes
 (Fig.~\ref{fig:Haloprofile1}), but they are indeed emptier in terms of total dark matter. This suggests that $f(R)$ 
gravity (or other modified gravities of the similar type) is unlikely be able to resolve the Local Void problem.
There are observations suggesting that the Local Void (within the radius of 1-8 Mpc from the center of the local group) seems 
too empty of galaxies, which may be a problem for the $\Lambda$CDM model \citep[e.g.][]{Peebles2001, Tikhonov2009, Peebles2010}, 
but see \citep{Tinker2009, Xie2014} for different views. There are speculations that a different model with 
more rapid emptying of voids and piling up of matter on its outskirts may help to resolve the tension \citep{Peebles2010}. 
At face value, it seems that $f(R)$ model is one of such models that has the required feature, as we have seen that 
the dark matter profiles in $f(R)$ are emptier and the void ridges are sharper. However, the more rapid halo formation 
in void regions in $f(R)$ models will perhaps make voids less empty of haloes. which is just the opposite to what is needed to resolve the tension. 
More detailed studies with simulations of better mass resolutions which are able to resolve lower mass haloes are needed to confirm this.}

\indent$\bullet$ We find that the halo number density profiles of voids in $f(R)$ are not distinguishable from those of GR. 
However, the dark matter density profiles associated with these voids are emptier of dark matter in $f(R)$ models than in GR; their over-density ridges, if any, are more prominent than those in GR. The latter result agrees with previous results based on the spherical expansion model of voids \citep{Clampitt2013}, which predicts that voids in $f(R)$ gravity are emptier and larger. Our results ring the alarm that void profiles can be very different depending on whether we are considering halo number or dark matter density profiles, even just in GR. This is not surprising from our perspective. Haloes or galaxies are biased tracers of the underlying dark matter density field. With relatively massive haloes whose linear bias is greater than unity, we do see the profiles of voids measured using haloes to be steeper than those of dark matter. This is essentially due to the fact that the amplitudes of the large-scale clustering of haloes are greater than those of dark matter. 
Note that this is different from the conclusion in \citep{SutterLavaux2014} that 
identical radial density profiles between galaxies and dark matter are recovered, though they have used a different void finder. i.e. ZOBOV.

\indent$\bullet$ Two different types of voids, void-in-cloud and void-in-void are clearly seen in either 
halo number density profiles or dark matter profiles, the former having sharp over dense ridges but not for the latter. 
This is consistent with the radial velocity profiles, which indicates a regime of mass inflow for void-in-clouds, but not for void-in-voids. 
This is also consistent with what has been found using ZOBOV for dark matter from simulations \citep{Hamaus2014,Paz2013}.
For this reason, we argue that in principle it is not possible to find a universal void profile with a single rescaling 
parameter, i.e. the void radius. A second parameter to characterize the height of the overdense ridge is necessary. 
Note that this is different from \citet{Nadathur2014} who find that voids in their simulated LRG catalogues are self-similar, 
but they have used ZOBOV as their void finder for mock and real LRGs, and applied further selections on their voids. 
Also, it should be noticed that the sizes of voids studied by \citet{Nadathur2014} are on the large size end of the void size
spectrum. These are usually of the void-in-void type according to our results and can indeed be described by
a radius scaling alone. We find that only smaller voids show a ridge and can be described as void-in-clouds. 
For them to be described by a density profile formula an extra parameter is needed \citep[e.g.][]{Paz2013}.

\indent$\bullet$ {The dark matter density profiles of voids can be measured using weak gravitational lensing. Observationally, this 
requires the combination of a galaxy redshift survey and a weak lensing (photometric redshift) survey over the same area of the sky. The idea of overlapping sky surveys has been promoted by the combination of redshift space distortion with lensing, which is another powerful way of constraining the linear growth and hence gravity \citep{McDonald2009, BC2011, Cai2012, Gaztanaga2012, dePutter2013, Kirk2013}. With this setting of surveys, voids will be identified using tracers like galaxies/haloes from the redshift survey. The shear signal of the background galaxies (from the deeper lensing survey) associated with the voids will be stacked around the void centres. We demonstrate that the lensing tangential shear profiles can be used to constrain $f(R)$ gravity. For a survey volume of about 1~(Gpc/$h$)$^3$, GR can be distinguished from F5 and F4 by  7 and 12$\sigma$ respectively, when including subvoids. Most of the signal comes from the edge of the voids where the density profiles are the steepest. The S/N is lower for larger voids as their abundance is much smaller and profiles less steep. We stress that the estimates of S/N are somewhat optimistic since we do not account for lensing shape noise and other systematics. Note that the S/N can be improved by a few times by employing the current BOSS survey or future EUCLID survey. We also found that including sub-voids is useful for increasing the S/N. Line-of-sight projection of large-scale structure degrades the S/N among models and may set limits on the constraint of $|f_{R0}|$. We find that F6 is not distinguishable from GR in terms of lensing of voids for this reason.} These S/N values increase when combining this statistics with the abundances of halo defined voids, making it
possible to increase the significance of the difference between F6 and GR to a $S/N\sim 2$. 

\indent$\bullet$ Admittedly, the steepening of dark matter void profiles in $f(R)$ model over GR may have some level of degeneracy with the 
increase of $\sigma_8$ in $\Lambda$CDM model. This can be checked with GR simulations of different $\sigma_8$. 
However, the halo void abundance in $f(R)$ being smaller than in GR is a unique feature that may be powerful to 
break this degeneracy. In this sense, it is important to combine measurements of void abundances and profiles.

\indent$\bullet$ Model differences in the velocity profiles are slightly greater than in the density profiles profiles, and these appear to be as strong when using the mass or tracers to measure velocities. This offers a good opportunity to constrain $f(R)$ gravity by studying voids in redshift space. This is particularly true for F4, which shows the strongest deviation 
from GR, while the constraints on F5 and F6 will be relatively weaker. A detailed study will be presented in a forthcoming paper. 
 
We caution that our study of void properties are based on using haloes as tracers. In principle, haloes are accessible through the observations of galaxy clusters and groups. Compared with galaxies, this observable suffers from the sparseness of samples and the relatively large uncertainties in determining the halo mass. Therefore, it may seem that galaxies are more direct observables to probe voids. However, the complexity of galaxy formation physics, which is likely to be different in $f(R)$ gravity from GR in a non-trivial manner, will make the definition of voids in different models even more complicated and the results less reliable. In contrast, haloes can be found observationally in different ways, such as using lensing, X-ray clusters and Sunyaev-Zel'dovich effects, which are less affected by the galaxy formation physics. Furthermore, most of our results concern only the number density of haloes rather than their masses, and so are less affected by the uncertainties caused by halo mass measurements (which can be worse in $f(R)$ gravity because haloes can have different dynamical and lensing masses, depending on their masses and environments).

{{Recently, there has been rapidly growing interest} in using cosmic voids to study cosmology or constrain gravity and dark energy models. In the two most well-studied categories of alternative gravity theories, the chameleon and Vainshtein types, the modification to GR is strongest in low-density regions, making voids a promising tool to constrain them. {In this paper, we have demonstrated for a specific example that this is indeed true}. Voids {as} a probe {for} cosmology {have} until very recently been considered as degraded by the ambiguities in the void definitions and void-finding algorithms. For example, \citet{Colberg2005} showed that different void finders do not agree even on the void abundances. Also, even applied on the same data set, different void-finding algorithms can output very different void density profiles \citep[see, for example,][where the latter group, using the tessellation code {\sc Zobov}, obtained significantly shallower void density profiles than what the former group found using a spherical under-density algorithm]{Pan2011,Nadathur2014}. However, as a void-finding algorithm is only a way to measure and quantify the distribution of dark matter, or its tracers such as galaxies and galaxy clusters, the details of the algorithm itself is less relevant as long as one uses the same algorithm to find voids in the mock universe (i.e., simulations) and the real one.} 

{Indeed, although often considered as a disadvantage, the ambiguity of defining and identifying voids can also have positive consequences: even if one void-finding algorithm is not sensitive to the modifications of gravity, others may well be, and by trying different algorithms one can hopefully find the optimal one to constrain a given type of gravity theory. For example, when determining the radius of a void, we have tried two different methods: (1) we divide particles around prospective void centres into a number of bins, find the bin at which the accumulative density first exceeds $\Delta_{\rm void}$, and let the void radius equal the radius of this bin; (2) not using the binning, but instead calculating the cumulative density every time when the void finder encounters another particle (or tracer), stopping when it first exceeds $\Delta_{\rm void}$, and taking the void radius as the distance between that particle and the void centre. Because of the sparseness of tracers in empty regions, we have found that the two methods can lead to different halo number density profiles, though the significant levels of deviation from GR and the dark matter density profiles (hence lensing signals) are stable. We will leave the study and comparison with other void-finding algorithms to future work.}

{Our results above have important implications {for} cosmological tests of gravity. It is often believed that cosmology can only place loose constraints on chameleon-type modified gravity theories, such as $f(R)$ gravity \citep[see, for example,][for a recent review]{lombriser2014}. Here, we see that void abundances, the stacking of void profiles and tangential shears and velocity profiles, can put constraints almost as strong as those from astrophysical tests, which suffer from bigger uncertainties. As such, it points out a new powerful probe which can potentially be applied to other types of gravity theories as well.}

\section*{Acknowledgments}

We thank Jiaxin Han, Marius Cautun, Pablo Arnalte-Mur, Shaun Cole, Carlos Frenk, Simon White, Rien van de Weygaert, Sergei Shandarin and John Peacock 
for helpful discussions and comments.
YC is supported by the Durham Junior Research Fellowship. {NP acknowledges
support by Fondecyt Regular No. 1110328 and BASAL PFB-06 CATA.} BL is supported by the Royal Astronomical Society and Durham University. 
YC and BL also acknowledge a grant with the RCUK reference ST/F001166/1.
NP acknowledges support by the European Commissions Framework Programme 7, through the Marie Curie International
Research Staff Exchange Scheme LACEGAL (PIRSES-GA-2010-269264).  

The simulations and part of data analysis for this paper were performed 
using the DiRAC Data Centric system at Durham University,
operated by the Institute for Computational Cosmology on behalf of the
STFC DiRAC HPC Facility (http://www.dirac.ac.uk). This equipment was funded by
BIS National E-infrastructure capital grant ST/K00042X/1, STFC capital
grant ST/H008519/1, and STFC DiRAC Operations grant ST/K003267/1 and
Durham University. DiRAC is part of the National E-Infrastructure.
Part of the analysis was done on the Geryon cluster at the Centre for Astro-Engineering UC, which
received recent funding from QUIMAL 130008 and Fondequip AIC-57. Access to the simulations used in 
this paper can be obtained from the authors.

\bibliography{voids}

\bibliographystyle{mn2e}

\section{Appendix}

This appendix contains the plot showing the enhancements in dark matter velocities due to the fifth force, using common void centres as found from the GR simulation. This should be read in the context of the discussion in the main paper.

\begin{figure*}
\begin{center}
\advance\leftskip -0.8cm
\scalebox{0.48}{
\includegraphics[angle=0]{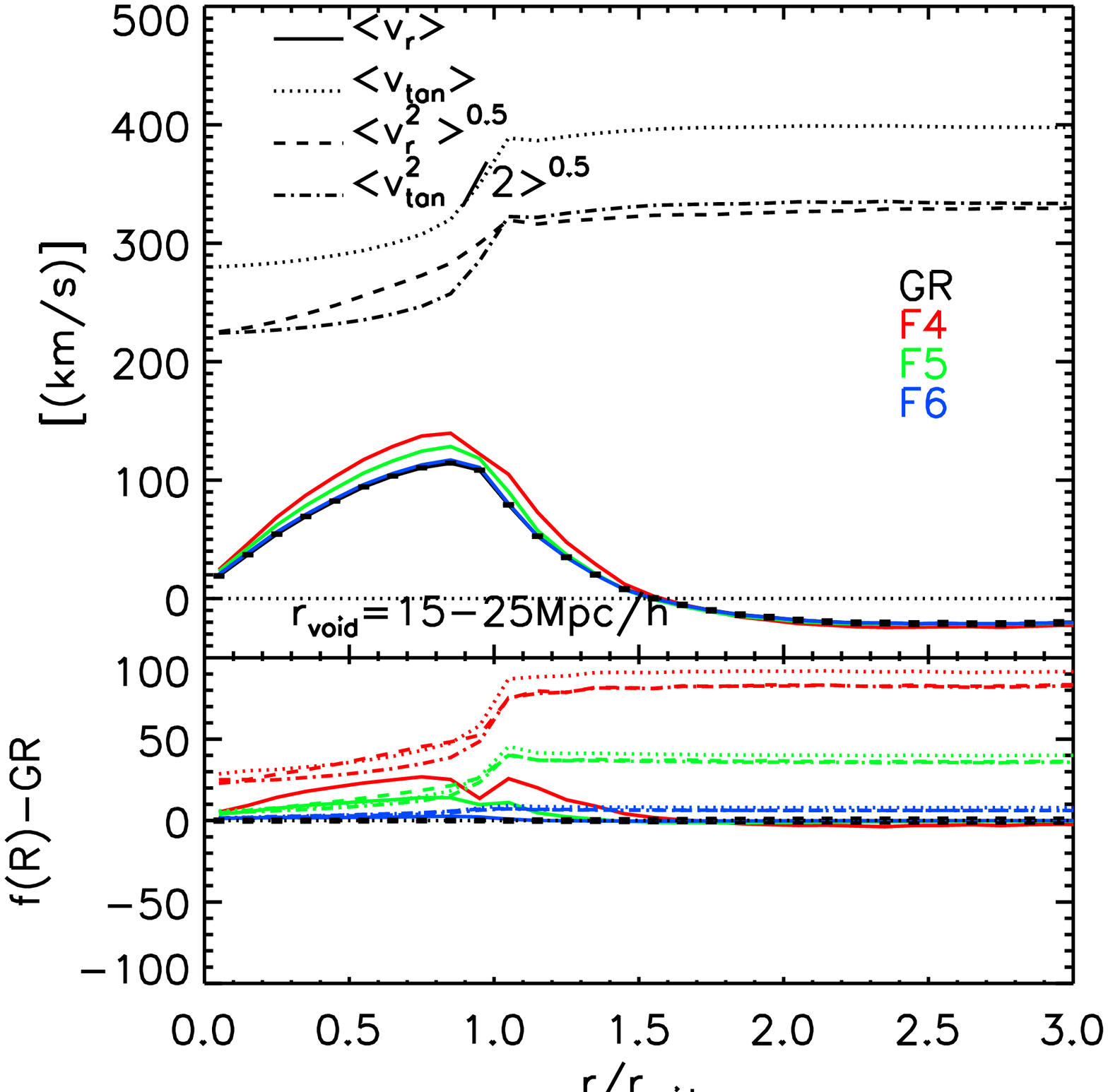}
\includegraphics[angle=0]{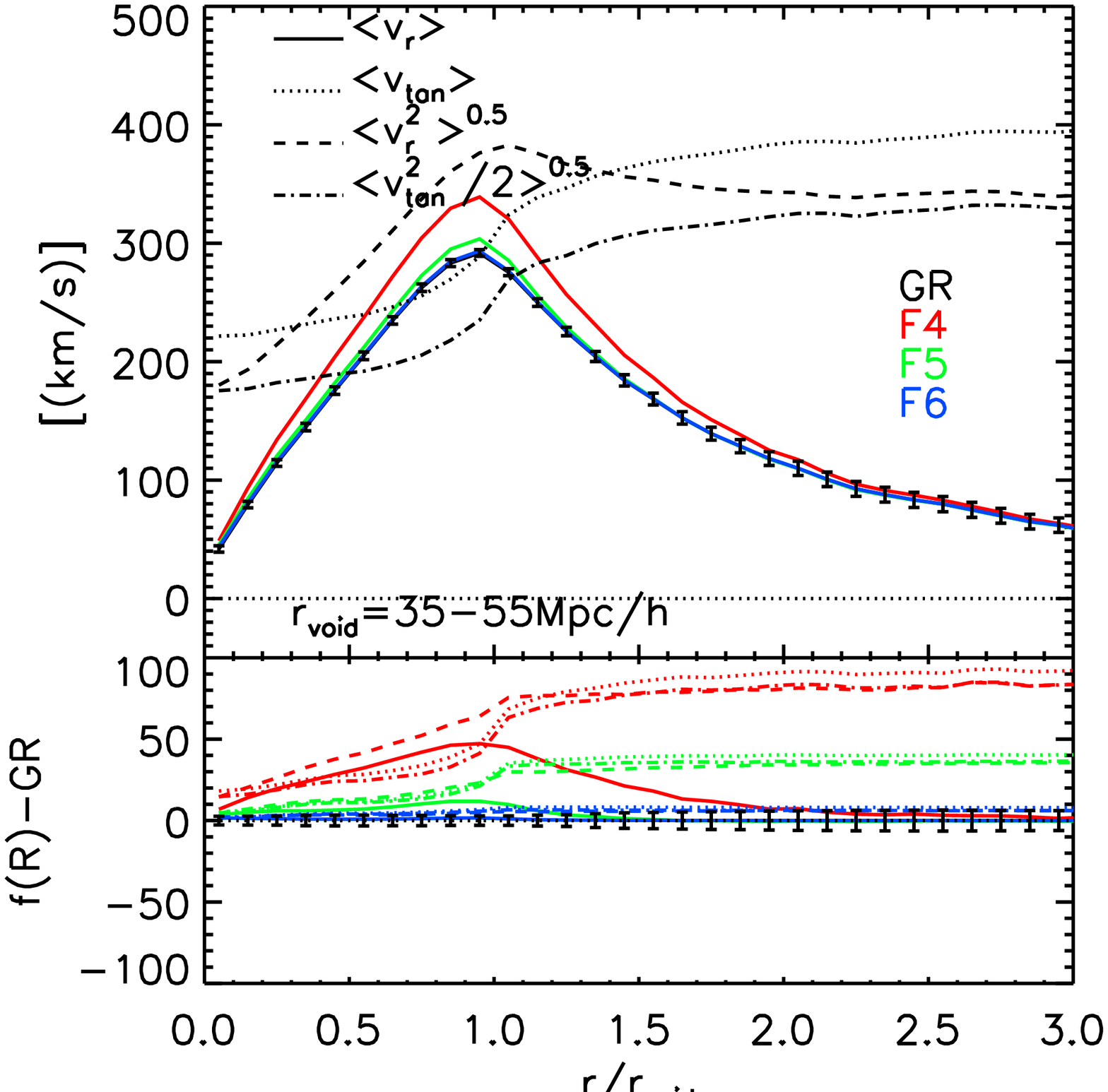}}
\caption{The same as Fig.~\ref{fig:Vprofile}, but showing results from using GR void centres to measure the profiles for all models. 
Top panels: different components of dark matter particle velocities {with} respect to the centre of 
voids defined using haloes for simulations of different models labelled in different colours in the legend. 
Solid lines -- the mean radial velocity profiles; dotted lines -- the mean tangential velocity profile;  
dashed lines -- the dispersion of the radial velocities; dash-dotted lines -- the dispersion of half of the tangential velocities. For simplicity, for velocity 
profiles of $f(R)$ models, only the radial velocity profiles are plotted (coloured solid lines). Their absolute differences with respect to GR are shown in the bottom panels. 
Error bars shown on the black line (GR) are the scatter about the mean for voids {with} $15 {\rm Mpc}/h<r_{\rm void}<25 {\rm Mpc}/h$ (left) and  
$35 {\rm Mpc}/h<r_{\rm void}<55 {\rm Mpc}/h$ (right) found in the 1 (Gpc/$h$)$^3$ simulation volume. }
\label{fig:Vprofile1}
\end{center}
\end{figure*}

\end{document}